\documentclass[12pt,preprint]{aastex}
\usepackage{amssymb}
\usepackage{amsmath}
\usepackage{graphicx}
\usepackage{subfigure}
\usepackage{wrapfig}
\usepackage[colorlinks,linkcolor=blue,anchorcolor=green,citecolor=blue]{hyperref}

\begin{document}

\def\func#1{\mathop{\rm #1}\nolimits}
\def\unit#1{\mathord{\thinspace\rm #1}}

\title{Probing the Birth of Post-merger Millisecond Magnetars by X-ray and
Gamma-ray Emission}
\author{Ling-Jun Wang\altaffilmark{1,2,3}, Zi-Gao Dai\altaffilmark{2,3},
Liang-Duan Liu\altaffilmark{2,3}, and Xue-Feng Wu\altaffilmark{4,5}}

\begin{abstract}
There is growing evidence that a stable magnetar could be formed from the
coalescence of double neutron stars. In previous papers, we investigated the
signature of formation of stable millisecond magnetars in radio and
optical/ultraviolet bands by assuming that the central rapidly rotating
magnetar deposits its rotational energy in the form of a relativistic
leptonized wind. We found that the optical transient PTF11agg could be the
first evidence for the formation of post-merger millisecond magnetars. To
enhance the probability of finding more evidence for the post-merger
magnetar formation, it is better to extend the observational channel to
other photon energy bands. In this paper we propose to search the signature
of post-merger magnetar formation in X-ray and especially gamma-ray bands.
We calculate the SSC emission of the reverse shock powered by post-merger
millisecond magnetars. We find that the SSC component peaks at $1\unit{GeV}$
in the spectral energy distribution and extends to $\gtrsim 10\unit{TeV}$
for typical parameters. These energy bands are quite suitable for \textit{%
Fermi}/LAT and CTA, which, with their current observational sensitivities,
can detect the SSC emission powered by post-merger magnetars up to $1\unit{%
Gpc}$. \textit{NuSTAR}, sensible in X-ray bands, can detect the formation of
post-merger millisecond magnetars at redshift $z\sim 1$. Future improvement
in sensitivity of CTA can also probe the birth of post-merger millisecond
magnetars at redshift $z\sim 1$. However, because of the $\gamma -\gamma $\
collisions, strong high-energy emission is clearly predicted only for ejecta
masses lower than $10^{-3}M_{\odot }$.
\end{abstract}



\affil{\altaffilmark{1}Key Laboratory of Space Astronomy and Technology,
National Astronomical Observatories, Chinese Academy of Sciences, Beijing
100012, China}

\affil{\altaffilmark{2}School of Astronomy and Space Science, Nanjing University, Nanjing 210093,
China; dzg@nju.edu.cn}

\affil{\altaffilmark{3}Key laboratory of Modern Astronomy and Astrophysics (Nanjing
University), Ministry of Education, Nanjing 210093, China}

\affil{\altaffilmark{4}Purple Mountain Observatory, Chinese Academy of
Sciences, Nanjing, 210008, China}

\affil{\altaffilmark{5}Joint Center for Particle Nuclear Physics and Cosmology of Purple Mountain
Observatory-Nanjing University, Chinese Academy of Sciences, Nanjing 210008, China}

\section{Introduction}

\label{sec:intro}

It is generally accepted that short gamma-ray bursts (SGRBs) result from
compact binary mergers, i.e. mergers of binary neutron stars (BNSs) or a
neutron star (NS) and a black hole (BH) %
\citep{paczynski86,eichler89,barthelmy05,fox05,gehrels05,rezzolla11}. Such a
picture of SGRBs can be confirmed by the upcoming next generation of
ground-based gravitational wave (GW) detectors %
\citep{Harry10,Somiya12,Acernese15,bartos13}.

The electromagnetic (EM) signals of compact binary mergers (NS-NS or NS-BH)
include SGRBs, kilonovae 
\citep[also known as
macronovae;][]{li98,kulkarni05,rosswog05,metzger10,roberts11,MetzgerBerger12,berger13,tanvir13,
Kasen15,Lippuner15}\footnote{%
There are other possibilities for the claimed kilonova associated with
GRB130603B \citep{takami14,Kisaka15,Kisaka16}}, radio afterglow %
\citep{nakar11,MetzgerBerger12,piran13,rosswog13} and possible X-ray
emission \citep{palenzuela13}.

The coalescence of BNSs appeals much attention because of its rich diversity
of possible outcome. In one popular scenario, the supramassive remnant of
BNS merger collapses into a black hole on a timescale of $\sim 200\unit{ms}$ %
\citep{rev-bin}. Hyperaccretion of a torus around the central black hole
could serve as the central engine for GRBs %
\citep{Popham99,Narayan01,Kohri02,Liu07,Liu15,Kawanaka13,Xue13,Song16}. In
addition, with the assumption of delayed fall-back disk accretion, the black
hole could also be the culprit for the X-ray bump and rebrightening
following some GRBs \citep{Wu13,Geng13,Yu15a}.

\cite{dai06} and \cite{zhang13}, on the other hand, put forward the
suggestion that the coalescence of BNSs could have another outcome, i.e. a
long-lasting or stable rapidly spinning magnetar. It is constantly
demonstrated that the magnetars could be responsible for the persistent
emission features of non-afterglow origin following a fraction of SGRBs %
\citep{dai06,rowlinson10,rowlinson13,Dai12,WangLJ13,Gompertz14,Gao15} and
the statistical properties of X-ray flares from some GRBs \citep{WangFY13}.
The formation of a stable magnetar following the merger of BNSs is confirmed
by recent simulations in numerical relativity %
\citep[e.g.][]{magnetar-sim,Giacomazzo15}.

To help identify the formation of stable post-merger magnetars, a line of
systematic research %
\citep{zhang13,gao13a,WangLJ13,Yu13,Yu15b,Metzger14,Wu14,wang15,Li16,Siegel16a,Siegel16b}
on the magnetar-aided EM signals has recently been carried out based on the
energy injection scenarios \citep{dai98a,dai98b,Zhang01,dai04}. One of the
main findings of these studies is that the optical/X-ray transients with a
GW association could be not associated with SGRBs. In previous papers %
\citep{WangLJ13,wang15}, under the assumption of Poynting flux of the
spinning-down magnetar becoming lepton dominated 
\citep[$e^+e^-$
pairs;][]{coroniti90,michel94,dai04,Yu07}, we studied the radio and
optical/ultraviolet emission based on the recently determined opacities of $%
r $-process material \citep{barnes13,kasen13}.

The basic picture of our model is schematized in Figure 1 of \cite{gao13a}
and Figure 1 of \cite{wang15}. The inspiral and final merger dynamically
eject a small amount of neutron-rich material with mass $M_{\mathrm{ej}%
}=10^{-4}$--$10^{-2}M_{\odot }$ and subrelativistic velocity $v=0.1$--$0.3c$ %
\citep{rezzolla10,goriely11,bauswein13,hotokezaka13,rosswog13}. The merger
also forms a torus with mass $\sim 10^{-2}M_{\odot }$ if the merger remnant
is a black hole \citep{Fernandez13}. By assuming a constant efficiency in
converting torus mass into GRB jet energy, \cite{Giacomazzo-etal13} find
that most of the tori have masses $\lesssim 10^{-2}M_{\odot }$. Neutrino
emission from the torus will drive a neutron-rich wind. An even larger
quantity of mass could be lost from the torus in a longer viscous time due
to outflows driven by viscous heating and recombination of free nuclei into $%
\alpha $-particles %
\citep[e.g.,][]{Metzger08,Metzger09,Lee09,Fernandez13,MetzgerFernandez14,Just15}%
. A long-lived magnetar may significantly enhance the mass loss from the
torus because of its long-lasting neutrino irradiation. \cite%
{MetzgerFernandez14} found that a fraction of $\sim 60\%$\ of the torus
mass, or equivalently $10^{-2}M_{\odot }$\ in mass, may be lost from the
torus provided that the merger remnant is a stable magnetar. The remaining
torus will be accreted in less than $\sim 1\unit{s}$ onto the (assumed)
remnant magnetar to power an SGRB %
\citep{ZhangD08,ZhangD09,ZhangD10,Giacomazzo-etal13}. We therefore expect
that the torus disappears after the emergence of SGRB. What left around the
magnetar after SGRB is the neutron-rich outflow with mass $\lesssim
10^{-2}M_{\odot }$.

The rapidly spinning massive neutron star builds up its magnetic field to
the magnetar level via differential rotation and begins to transfer its
rotational energy to the ejecta in the form of Poynting flux $L_{\mathrm{sd}%
}=L_{\mathrm{sd},0}/\left( 1+t/T_{\mathrm{sd}}\right) ^{2}$, where the
spin-down luminosity is $L_{\mathrm{sd},0}=10^{47}\unit{erg}\unit{s}%
^{-1}P_{0,-3}^{-4}B_{p,14}^{2}R_{6}^{6}$, the spin-down timescale is $T_{ 
\mathrm{sd}}=2\times 10^{5}\unit{s}R_{6}^{-6}B_{p,14}^{-2}P_{0,-3}^{2}$, $%
P_{0}$ is the initial rotation period of magnetar, $B_{p}$ is the dipole
magnetic field of the magnetar, and $R$ is the neutron star radius.
Throughout this paper we adopt the usual convention $Q=10^{n}Q_{n}$. The
Poynting flux eventually becomes $e^{\pm }$-dominated and the transition
from Poynting-flux-dominated wind to $e^{\pm }$-dominated wind could be
abrupt \citep{Aharonian12}. The relativistic magnetar wind is braked by the
slow ejecta as discussed above and therefore a reverse shock develops. The
accelerated ejecta drive into the ambient interstellar medium and a forward
shock develops. The ejecta can be treated as a thin shell \citep{wang15}
despite the effects of radioactive heating \citep{rosswog14}.

We showed that the optical transient PTF11agg \citep{cenko13} discovered by
the wide-field survey Palomar Transient Factory can be nicely interpreted as
the reverse shock synchrotron emission of the leptonized relativistic wind
powered by a post-merger millisecond magnetar \citep{WangLJ13}. In the paper
that follows \citep{wang15}, we systematically investigated all three cases
that were studied by \cite{gao13a}. We found that the very early broadband
emission at different wavelengths depends on the properties of the $r$%
-process material and an ionization breakout is expected based on current
knowledge of the $r$-process material.

The electrons in the reverse shock are ultra-relativistic so that inverse
Compton (IC) would have an influential effect on the radiative process of
electrons. The intense flux of UV/X-ray synchrotron photons produced at the
reverse shock serves as seed photons for the relativistic $e^{\pm }$ pairs
to inverse Compton scatter them to ultrahigh energy (UHE). In Section \ref%
{sec:gamma-gamma-collision} we show that the thermal and synchrotron photons
in the reverse shock could influence the high-energy IC radiation,
especially during the early time for the highest energy photons. We find
that the IC emission during the magnetar spin-down will be completely
attenuated if the ejecta mass $\gtrsim 3\times 10^{-3}M_{\odot }$.

Based on this result, in this paper, we mainly restrict our attention to the
IC emission for Case I, i.e., the low ejecta mass case ($M_{\mathrm{ej}%
}<10^{-3}M_{\odot }$) studied by \cite{WangLJ13}. Our knowledge about the
electromagnetic emission and composition of a newborn magnetar wind comes
from the observation and modeling of PWNe. So in Section \ref{sec:Background}
we clarify the theoretical and observational background of PWNe with an aim
to justify our treatment that follows, and leave the open questions in
Section \ref{sec:discuss}. Section \ref{sec:syn-IC} is divided into two
subsections. In Section \ref{sec:syn} we analytically evaluate the effects
of IC on synchrotron emission, while in Section \ref%
{sec:gamma-gamma-collision} we analytically estimate the attenuation of IC
photons by the low-energy synchrotron and thermal photons. In Section \ref%
{sec:ic} we propose to detect the birth of post-merger millisecond magnetars
by gamma-ray observations. It is found that the emission at the energy band
of $1\unit{GeV}$ and $100\unit{GeV}$, suitable for Fermi Large Area
Telescope \citep[Fermi/LAT,][]{Atwood09} and Cherenkov Telescope Array %
\citep[CTA,][]{Actis11} respectively, can be detected up to distances $\sim 1%
\unit{Gpc}$ $\left( z\approx 0.2\right) $ with their current detection
sensitivities for a typical millisecond magnetar. We conclude our main
findings in Section \ref{sec:conclusion}.

\section{Theoretical and Observational Background}

\label{sec:Background}

Theoretically, several channels may give birth to millisecond magnetars. One
channel to form millisecond magnetars is the double neutron star mergers %
\citep{dai98a,dai98b}, viz. the case investigated in this paper. The second
channel is the collapse of massive stars, whereby the formation of
millisecond magnetars manifests themselves by the emergence of superluminous
supernovae (SLSNe) 
\citep{Kasen10,Woosley10, Chatzopoulos12,Inserra13,Nicholl14,
Papadopoulos15,WangLiu15,Wang15b,Wang16,Dai16} and luminous SNe %
\citep{Greiner15,Wang15c}. The third channel is the merger of an NS and a
white dwarf \citep[WD;][]{Metzger12} with an ejecta mass of $0.3-1M_{\odot }$%
. In the latter two cases the ejecta are so massive that the high energy
emission cannot be observed immediately following the formation of
millisecond magnetars.

The fourth possible channel to form a millisecond magnetar is the
accretion-induced collapse (AIC) of a WD \citep{Canal76,Ergma76}.
Radiation-hydrodynamics simulations suggest that a total mass of a few times 
$10^{-3}M_{\odot }$ be ejected during the collapse of the WD to NS %
\citep{Dessart06}. Such a mass is high enough to block the gamma rays
emanated from a reverse shock during the early spin-down time of the nascent
millisecond magnetar. In addition, whether or not the WD will collapse into
an NS or undergo a thermonuclear explosion as a Type Ia supernova in an AIC
depends on the mass accretion rate, the WD mass, and the WD composition %
\citep{Nomoto82,Nomoto84,Nomoto91}. The collapse rate of a WD into an NS
could be very low and the ejecta composition is also quite different from
that coming from the compact binary mergers \citep{Fryer99}. Based on these
considerations, we will restrict ourselves to the magnetars formed during
the coalescence of BNSs in this paper.

The rotational energy of the magnetar is converted as Poynting flux within
the light cylinder, as can be inferred from the large value of the
magnetization parameter $\sigma $, i.e. the ratio of magnetic to particle
kinetic energy flux of the flow \citep{Michel82,Gaensler06,Hester08}.
However, to reproduce the observational properties of pulsar wind nebulae
(PWNe) the magnetar wind must become particle-dominated upstream of the
reverse shock, i.e. the termination shock of the synchrotron PWNe %
\citep{Rees74,Kennel84a,Kennel84b,Begelman92}. The transition mechanism from
a high $\sigma $ wind to a low $\sigma $ one, known as the $\sigma $%
-problem, is as yet poorly understood despite circa five-decade
investigations \citep{Kargaltsev15}.

One idea that has received a great deal of attention in the past decades is
the magnetic reconnection triggered by the annihilation of the alternating
magnetic fields (striped wind) near the equatorial plane, generated by an
obliquely rotating pulsar, between the light cylinder and the termination
shock 
\citep{Michel82,michel94,coroniti90,Lyubarsky01,Kirk03,Lyubarsky03,
Lyubarsky05,Lyubarsky10a,Lyubarsky10b,Petri07,Petri08,Arons12,Hoshino12}. 1D
spherical and 2D axisymmetric MHD models \citep{Atoyan96,Volpi08,Olmi14}
persistently require low value of $\sigma $ upstream of the termination
shock. Recent 3D simulations of plasma flow \citep{Mizuno11,Porth14} suggest
instead that the value of $\sigma $ upstream of the termination shock need
not to be as low as previously thought, alleviating the $\sigma $-problem.

What makes things even more involved is the abrupt acceleration of the cold
ultrarelativistic wind at $20-50R_{\mathrm{LC}}$ \citep{Aharonian12} as
inferred under the plausible assumption that the very high energy emission
from Crab nebula results from the inverse Compton scattering of pulsed X-ray
photons. This fact indicates that the pulsar wind becomes lepton-dominated
and $\sigma \ll 1$ far within the termination shock, challenging current
theoretical understanding of the pulsar wind magnetization. On the
theoretical side, 1D spherically symmetric MHD models found that the
distance of the termination shock from the Crab pulsar could be reproduced
only if $\sigma \sim 0.003$ upstream of the shock \citep{Kennel84a,Kennel84b}%
. 2D simulations indicate that $\sigma \sim 0.02$ works well in accounting
for the location of the termination shock and the nebular morphology %
\citep{Buhler14}. This value is further raised in recent 3D simulations, $%
\sigma >1$ \citep{Mizuno11,Porth14}.

But the price in achieving this goal is that the dipolar jet is much weaker
than observed \citep{Porth14}. This argues for an efficient dissipation to
convert magnetic energy into dipolar jet energy, in accord with the
discovery of abrupt acceleration of the cold wind, viz. a small $\sigma $.
For the newborn millisecond magnetars considered in this paper, it is
expected that $\sigma $ is even smaller because of the strong dipolar
magnetic field and rapid spinning. In the striped wind model, the faster the
pulsar is spinning, the shorter the stripe wavelength. In addition, the
stronger the dipolar magnetic field, the easier the magnetic reconnection to
occur. These two aspects make the dissipation of Poynting flux to particle
flux more efficient for millisecond magnetars.

If $\sigma $ is low enough, i.e. $\sigma \lesssim 10^{-3}$ \citep{Sironi11a}%
, the two-stream Weibel instability \citep{Fried59,Weibel59,Medvedev99} in
the plasma will set in and stochastic magnetic field will grow. If, on the
other hand, $\sigma $ is high enough, the Weibel instability will be
suppressed and the magnetic field in the shocked wind will be dominant due
to the compression of the unshocked wind magnetic field \citep{Zhang05}. We
cannot exactly evaluate $\sigma $ for a millisecond magnetar based on
current theoretical understanding of the magnetic dissipation in
relativistic wind. As we argued above, the value of $\sigma$ could be much
small for a millisecond magnetar. On the observational aspect, the electrons
accelerated behind the termination shock of the pulsar wind have a power-law
distribution with an index $p=2.1-2.8$ \citep{Kargaltsev15}, as expected
from the relativistic collisionless shock acceleration %
\citep{Achterberg01,Sironi09,Sironi11a,Buhler14}. Furthermore, \cite%
{WangLJ13} found that the optical transient PTF11agg can be nicely
interpreted as an unmagnetized reverse shock emission powered by a
millisecond magnetar. This seems to indicate that the magnetization
parameter of the wind blown out by the newborn millisecond magnetar powering
PTF11agg is quite low, i.e. $\sigma \lesssim 10^{-3}$. In summary, we argue
that for millisecond magnetar, $\sigma $ is so low that the Weibel
instability could grow and the electrons in the reverse shock have a
power-law distribution with index $2<p<3$.

The magnetic field in the striped wind cannot completely annihilate into
electron pairs because the field lines are poloidal near the axis rather
than alternating as they are near the equatorial plane. This seems to pose a
concern for our assumption that the wind is dominated by leptons. In fact,
however, the pulsar wind is highly anisotropic so that the Poynting flux
varies with the polar angle as $\sin ^{2}\theta $ %
\citep{Michel73,Bogovalov99,Komissarov13,Kargaltsev15}, i.e. the Poynting
flux decreases monotonically from the equatorial plane to the polar axis.
Numerical simulations show an even more anisotropic distribution that varies
as $\sin ^{4}\theta $ \citep{Tchekhovskoy13}. Therefore the rotational
energy of the pulsar is carried away predominantly along the equatorial
plane where the magnetic energy is transferred into electron pairs
efficiently and our treatment is justified.\footnote{%
In some cases strong dissipation is observed near the rotation axis, where
the column of baryonic material is likely to be minimized. However, this
dissipation along the rotation axis is usually negligibly small compared to
the spin-down power of the pulsar.}

The particle Lorentz factor of the pulsar wind usually takes on the values
of $\gamma _{4}\sim 10^{4}-10^{7}$ %
\citep{Kennel84a,Kennel84b,atoyan99,Michel99,Fang10,Bucciantini11,Tanaka11,WangLJ13}%
. In this paper we calculate the IC emission by taking two typical values of 
$\gamma _{4}=10^{4}$ and $10^{6}$.

The neutron star is also an astrophysical thermal emitter \citep{Potekhin15}
whose thermal photons from its surface should in principle be taken into
consideration. However, because the newborn neutron star cools so rapidly
that when the IC photons from the reverse shock manage to penetrate the
ejecta, the thermal photons from the neutron star surface become negligible
compared with the synchrotron photons. Consequently we ignore thermal
emission from the neutron star surface.

However, the thermal radiation comes along with the pairs cannot be ignored.
In this paper, to be able to tackle the problem in an analytical way and
also due to the uncertainty of modeling the thermal radiation accompanied
with electron pairs, we assume that the thermal radiation generated along
with the pairs takes the form $L_{\mathrm{th}}\propto L_{\mathrm{sd}}$. This
assumption is reasonable because thermalization occurs predominantly near
the light cylinder and is independent of the ejecta and reverse shock.
Figure 2 of \cite{wang15} shows that the synchrotron luminosity $L_{\mathrm{%
syn}}$ during the spin-down time of the magnetar is nearly a constant. We
therefore assume that 
\begin{equation}
L_{\mathrm{th}}=\eta _{\mathrm{th}}L_{\mathrm{syn}}  \label{eq:Lth-Lsyn}
\end{equation}%
when $t<T_{\mathrm{sd}}$. For $t>T_{\mathrm{sd}}$ the rotational energy of
the magnetar is exhausted and we assume $L_{\mathrm{th}}=0$.\footnote{%
To make the analytic calculations affordable, we simplify the problem by
assuming that the spin-down power shuts off at the spin-down time, as done
in the literature \citep{gao13a,WangLJ13,wang15}. In alignment with this
simplification, here we assume that $\eta _{\mathrm{th}}=0$ at spin-down
time. Nevertheless, this simplification will not loss important physics
because the thermal photons only affect the IC photons at the cut-off
energy, as shown in Section \ref{sec:gamma-gamma-collision}.} We caution
that although $L_{\mathrm{syn}}$ is constant during the spin-down of the
magnetar, but $L_{\mathrm{syn}}$ depends not only on $L_{\mathrm{sd}}$ but
also on the ejecta mass. As a result, $\eta _{\mathrm{th}}$ is not a
universal constant and varies from case to case. In the following
calculations, we assume $\eta _{\mathrm{th}}=0.2$.

\section{High-energy radiation by IC}

\label{sec:syn-IC}

\subsection{Effects of IC on Synchrotron Emission}

\label{sec:syn}

The inverse Compton emission of GRB afterglows has been studied and detected
for many GRBs %
\citep{Panaitescu98,Wei98,Panaitescu00,Wang01a,Wang01b,Harrison01,Sari01,Nakar09,Wang10,Liu13,WangK13}%
. It is expected that the IC component of the reverse shock emission in our
model is prominent because of the ultrarelativistic nature of the shocked
electrons. The IC scattering will boost the cooling of radiative electrons
so that the total power of one relativistic electron is%
\begin{equation}
P(\gamma _{e})=P_{\mathrm{syn}}\left( \gamma _{e}\right) \left( 1+Y\right) ,
\end{equation}%
where%
\begin{equation}
P_{\mathrm{syn}}\left( \gamma _{e}\right) =\frac{4}{3}\sigma _{T}c\gamma
^{2}\beta _{e}^{2}\gamma _{e}^{2}\frac{B^{2}}{8\pi },
\end{equation}%
$Y$ is the Compton $Y$ parameter and other parameters have their usual
meanings \citep{Sari98}. As a result, the electron cooling Lorentz factor is
modified as%
\begin{equation}
\gamma _{c}=\frac{6\pi m_{e}c}{\sigma _{T}\gamma B^{2}t\left( 1+Y\right) }.
\label{eq:cooling-Lorentz-factor-IC}
\end{equation}

In determining the ejecta dynamics, we have adopted two different equations,
i.e., Equations $\left( \ref{eq:PTF11agg-dyn}\right) $ and $\left( \ref%
{eq:dynamics-RS-work}\right)$ in \cite{WangLJ13} and \cite{wang15}, as
follows, 
\begin{equation}
L_{0}\min \left( t,T_{\mathrm{sd}}\right) =\left( \gamma -\gamma _{\mathrm{ej%
},0}\right) M_{\mathrm{ej}}c^{2}+2\left( \gamma ^{2}-1\right) M_{\mathrm{sw}%
}c^{2},  \label{eq:PTF11agg-dyn}
\end{equation}%
from the energy conservation of the whole system, while 
\begin{equation}
\frac{d\gamma }{dt}=\frac{\xi L_{\mathrm{sd}}+L_{\mathrm{rd}}-L_{\mathrm{ej,}%
e}-\gamma \mathcal{D}\left( \frac{dE_{3}^{\prime }}{dt^{\prime }}+\frac{dE_{%
\mathrm{ej,int}}^{\prime }}{dt^{\prime }}\right) -\left( {\gamma }%
^{2}-1\right) {c}^{2}\left( \frac{dM_{\mathrm{sw}}}{dt}\right) }{%
E_{3}^{\prime }+M_{\mathrm{ej}}c^{2}+E_{\mathrm{ej,int}}^{\prime }+2\gamma
M_{\mathrm{sw}}c^{2}},  \label{eq:dynamics-RS-work}
\end{equation}%
for the dynamics in the differential way, where $L_{0}=\xi L_{\mathrm{sd}}$,
with $\xi $ the fraction of spin-down power of the magnetar that is caught
by the ejecta, $L_{\mathrm{sd}}$ the spin-down power of the magnetar, $T_{%
\mathrm{sd}}$ the spin-down timescale, and $\gamma_{\mathrm{ej},0}$ the
initial Lorentz factor of the ejecta, $L_{\mathrm{rd}}$ the radioactive
heating power due to the $r$-process material, $L_{\mathrm{ej,}e}$ the
thermal luminosity of the ejecta, $\mathcal{D}$ the Doppler factor, $%
E_{3}^{\prime }$ and $E_{\mathrm{ej,int}}^{\prime }$ the respective energy
in Region 3 and in ejecta in the comoving frame. The four regions, i.e.
Region 1 to 4, are defined as: unshocked medium (Region 1), shocked medium
(Region 2), shocked wind (Region 3), and unshocked magnetar wind (Region 4),
see Figure 1 in \cite{wang15}.

Equation $\left(\ref{eq:PTF11agg-dyn}\right) $ originates from Equation $%
\left( 1\right) $ in \cite{gao13a} and the fact that the energy contained in
forward shock and reverse shock is comparable \citep{blandford76,WangLJ13}.
Equation $\left( \ref{eq:dynamics-RS-work}\right) $, on the other hand, is
determined by accounting for all energy in different zones %
\citep{dai04,Yu13,wang15}.

In this section we aim to evaluate the effects of IC on synchrotron spectra
and light curves in an analytical manner. To this end it is affordable to
adopt Equation $\left( \ref{eq:PTF11agg-dyn}\right) $ rather than $\left( %
\ref{eq:dynamics-RS-work}\right) $ for the dynamical evolution of the ejecta
because \cite{wang15} showed that Equation $\left( \ref{eq:PTF11agg-dyn}%
\right) $ is quantitatively equivalent to Equation $\left( \ref%
{eq:dynamics-RS-work}\right) $.

As shown in \cite{WangLJ13}, to analytically calculate the ejecta dynamics
and the reverse shock light curves, several timescales have been defined %
\citep[see Figures 1 and 2 in ][for reference.]{WangLJ13}: the transition
times between Newtonian dynamics and relativistic dynamics $T_{N1}$ and $%
T_{N2}$, the ejecta deceleration time $T_{\mathrm{dec}}$, the spin-down time
of the magnetar $T_{\mathrm{sd}}$, and the time, $T_{\mathrm{ct}}$, i.e. the
time at which electron cooling factor $\gamma _{c}$ begins to deviate from
unity. The reason for defining $T_{\mathrm{ct}}$ is that the electrons in
the reverse shock cool so efficiently at beginning that $\gamma _{c}=1$
before $T_{\mathrm{ct}} $. There are two other timescales that are important
for the calculation of reverse shock light curves: $T_{ac}$ and $T_{mc}$,
the respective crossing time of cooling frequency $\nu _{c}$ with
synchrotron self-absorption frequency $\nu _{a}$ and the typical frequency $%
\nu _{m}$.

IC scattering affects the cooling of electrons so we shall expect that the
timescales $T_{\mathrm{ct}}$, $T_{ac}$ and $T_{mc}$ would be modified when
IC is taken into account. The Compton $Y$ parameter is given by %
\citep{Sari01} 
\begin{equation}
Y=\frac{1}{2}\left( \sqrt{1+4\eta \frac{\epsilon _{e}}{\bar{\epsilon}_{B}}}%
-1\right) ,  \label{eq:exactx}
\end{equation}%
which can be simplified as%
\begin{equation}
Y=\left\{ 
\begin{array}{ll}
\frac{\eta \epsilon _{e}}{\bar{\epsilon}_{B}}, & \frac{\eta \epsilon _{e}}{%
\bar{\epsilon}_{B}}\ll 1, \\ 
\left( \frac{\eta \epsilon _{e}}{\bar{\epsilon}_{B}}\right) ^{1/2}, & \frac{%
\eta \epsilon _{e}}{\bar{\epsilon}_{B}}\gg 1,%
\end{array}%
\right.  \label{eq:approxx}
\end{equation}%
in the limiting cases. Here $\eta $ is the fraction of the electron energy
that is radiated away by synchrotron and IC emission. Compared with \cite%
{Sari01}, here we introduce a new parameter%
\begin{equation}
\bar{\epsilon}_{B}=\left\{ 
\begin{array}{lc}
\epsilon _{B}\left( 1+\eta _{\mathrm{th}}\right) , & t<T_{\mathrm{sd}} \\ 
\epsilon _{B}, & t>T_{\mathrm{sd}}%
\end{array}%
\right.
\end{equation}%
according to the discussion given in Section \ref{sec:Background}. $\epsilon
_{e}$ and $\epsilon _{B}$ are the fraction of total energy going into
electrons and magnetic field, respectively. In the following analytical
calculations, we usually take the approximation $\bar{\epsilon}_{B}\approx
\epsilon _{B}$.

Before $T_{mc}$, i.e. the transition time between fast cooling and slow
cooling, the electrons in the reverse shock are in the fast cooling regime
so that $\eta =1$. As determined by \cite{WangLJ13}, $\epsilon _{B}\simeq
0.1 $ so that $\epsilon _{e}/\epsilon _{B}\simeq 9\gg 1$. By Equation $%
\left( \ref{eq:approxx}\right) $ and the definition of $T_{\mathrm{ct}}$ %
\citep{WangLJ13}%
\begin{equation}
T_{\mathrm{ct}}=\frac{3\pi m_{e}c}{\sigma _{T}\gamma B_{3}^{2}\left(
1+Y\right) },
\end{equation}%
we find

\begin{equation}
T_{\mathrm{ct}}=7.3\times 10^{-2}\unit{days}L_{0,47}^{-2/3}M_{\mathrm{ej}%
,-4}^{5/6}\epsilon _{e}^{1/12}\epsilon _{B,-1}^{1/12}.
\end{equation}%
In this approximation, i.e. $Y\gg 1$, the other two timescales can also be
determined%
\begin{equation}
T_{ac}=0.141\unit{days}L_{0,47}^{-\left( 8p+25\right) /2\left( 6p+19\right)
}M_{\mathrm{ej},-4}^{\left( 5p+16\right) /\left( 6p+19\right) }\text{$%
\epsilon $}_{B,-1}^{\left( p+2\right) /2\left( 6p+19\right) }\text{$\epsilon 
$}_{e}^{\left( p+3\right) /2\left( 6p+19\right) }\text{$\gamma $}%
_{4,4}^{-1/\left( 6p+19\right) },
\end{equation}%
\begin{equation}
T_{mc}=0.144\unit{days}L_{0,47}^{-5/7}M_{\mathrm{ej},-4}^{6/7}\text{$%
\epsilon $}_{B,-1}^{1/14}\text{$\epsilon $}_{e}^{3/14}\text{$\gamma $}%
_{4,4}^{1/7}.
\end{equation}%
The above numerical values are evaluated for $p=2.2$, a value preferred in 
\cite{WangLJ13}. Throughout this paper we evaluate all reverse shock
parameters at this value of $p$.

Before $T_{mc}$ the effect of IC on the cooling frequency $\nu _{c}$ is just
to modify it by a constant factor so that the temporal scaling indices of $%
\nu _{c}$ is not amended. But after $T_{mc}$, i.e., when the electrons are
in the slow cooling regime, not only the exact value of $\nu _{c}$, but also
its temporal scaling indices will be changed if IC dominates the electron
cooling.

\cite{WangLJ13} showed that $T_{\mathrm{dec}}$ is given by 
\citep[see
also][]{gao13a}%
\begin{equation}
T_{\mathrm{dec}}=0.28\unit{days}L_{0,47}^{-7/10}M_{\mathrm{ej}%
,-4}^{4/5}n^{-1/10},  \label{eq:T_dec}
\end{equation}%
which is usually larger than $T_{mc}$. Now let us calculate the evolution of 
$\nu _{c}$ in the time period $T_{mc}<t<T_{\mathrm{dec}}$. Before $T_{mc}$
the electrons are in the fast cooling regime and IC dominates their cooling.
So we expect that IC will continue to dominate the cooling of electrons for
a while.

In the time period $T_{mc}<t<T_{\mathrm{dec}}$, if IC cooling is ignored, we
have $\nu _{c}\propto t^{9}$ and $\nu _{m}\propto t^{-5}$ \citep{WangLJ13}
so that $\nu _{c}/\nu _{m}\propto t^{14}$. When IC cooling is taken into
account, in the slow IC-dominated stage, we have \citep{Sari01}%
\begin{equation}
\nu _{c}/\nu _{m}=\left( t/T_{mc}\right) ^{14}Y^{-2}.
\end{equation}%
Please note that, upon taking into account IC emission, $T_{mc}$ is no
longer the time when $\nu _{c}=\nu _{m}$. Here $T_{mc}$ is similar to the
variable $t_{0}^{\mathrm{IC}}$ defined by \cite{Sari01}. Substitution of
Equation $\left( \ref{eq:approxx}\right) $ into the above equation in the
case of $\eta \epsilon _{e}/\epsilon _{B}\gg 1$ and knowing that%
\begin{equation}
\eta =\left( \gamma _{c}/\gamma _{m}\right) ^{-\left( p-2\right) },
\end{equation}%
for slow cooling, we have%
\begin{equation}
Y\simeq \left[ \left( \frac{t}{T_{mc}}\right) ^{14}\frac{1}{Y^{2}}\right]
^{-\left( p-2\right) /4}\sqrt{\frac{\epsilon _{e}}{\epsilon _{B}}},
\end{equation}%
which gives%
\begin{equation}
Y\simeq \left( \frac{\epsilon _{e}}{\epsilon _{B}}\right) ^{1/(4-p)}\left( 
\frac{t}{T_{mc}}\right) ^{-7(p-2)/(4-p)}.  \label{eq:Y-T_mc-T_dec}
\end{equation}%
We finally arrive at for the interval $T_{mc}<t<T_{\mathrm{dec}}$:%
\begin{equation}
\frac{\nu _{c}}{\nu _{m}}=\left( \frac{t}{T_{mc}}\right) ^{28/\left(
4-p\right) }\left( \frac{\epsilon _{e}}{\epsilon _{B}}\right) ^{-2/\left(
4-p\right) }.
\end{equation}

IC emission will not dominate the electron cooling when $Y=1$, which occurs
at%
\begin{equation}
t^{\mathrm{IC}}=T_{mc}\left( \frac{\epsilon _{e}}{\epsilon _{B}}\right)
^{1/7\left( p-2\right) }\simeq 0.69\unit{days}.
\end{equation}%
This time is usually later than $T_{\mathrm{dec}}$. Consequently, IC
emission usually dominates the electron cooling in the entire stage $%
T_{mc}<t<T_{\mathrm{dec}}$. The IC emission will gradually become less
dominant for electron cooling when the temporal scaling indices of $\nu
_{c}/\nu _{m}$ is positive. However, inspection of Table 1 of \cite{WangLJ13}
indicates that the temporal scaling indices of $\nu _{c}/\nu _{m}$ is
non-positive when $t>T_{\mathrm{dec}}$. In other words, the IC emission
becomes progressively dominant for electron cooling in the stage $t>T_{%
\mathrm{dec}}$.

The cooling Lorentz factor in the slow IC-dominated regime is given by

\begin{equation}
\gamma _{c}=\left[ \frac{6\pi m_{e}c}{\sigma _{T}\gamma B^{2}t}\left( \frac{%
\epsilon _{B}}{\epsilon _{e}}\right) ^{1/2}\frac{1}{\gamma _{m}^{\left(
p-2\right) /2}}\right] ^{2/\left( 4-p\right) },
\end{equation}%
from which we can analytically calculate $\nu _{c}$ for $t>T_{\mathrm{dec}}$%
. The calculated temporal scaling indices of $\nu _{c}$ are given in Table %
\ref{tbl:indices-nu}. It is worth mentioning that because $p$ is very close
to 2, the scaling indices of $\nu _{c}$ given in Table \ref{tbl:indices-nu}
are actually nearly identical to that given in Table 1 of \cite{WangLJ13}.
As a result, it is usually accurate enough to use Table 1 of \cite{WangLJ13}
to assess the time evolution of the various synchrotron quantities.

\begin{figure}[tbph]
\centering\includegraphics[width=1.0\textwidth,angle=0]{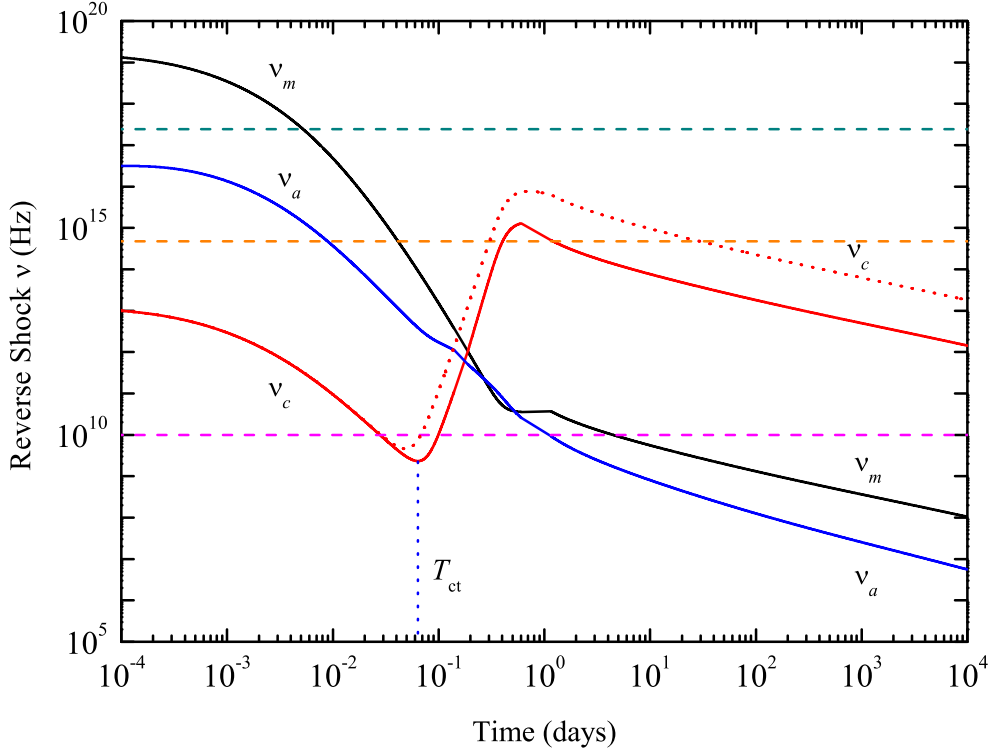}
\caption{The evolution of $\protect\nu _{c}$, $\protect\nu _{m}$, and $%
\protect\nu _{a}$ (solid lines) when IC cooling is taken into account. For
comparison, $\protect\nu _{c}$ without IC cooling is shown as the dotted
line. In this numerical calculation, we adopt Equation $\left( \protect\ref%
{eq:dynamics-RS-work}\right) $ to calculate the ejecta dynamics. The three
horizontal dashed lines mark the X-ray, optical (R) and radio (10 GHz)
bands, respectively. The vertical dotted line indicates the time $T_{\mathrm{%
ct}}$.}
\label{fig:cooling-nu}
\end{figure}

To quantitatively appreciate the effects of IC on the synchrotron emission,
in Figure \ref{fig:cooling-nu} we compare the evolution of synchrotron
characteristic frequencies with/without IC taken into account. In this
numerical calculation, we adopt Equation $\left( \ref{eq:dynamics-RS-work}%
\right) $ to determine the ejecta dynamics. Before $T_{mc}$, IC cooling is
to introduce an extra constant factor $\left( \epsilon _{e}/\epsilon
_{B}\right) ^{1/2}$ for $\nu _{c}$. As a result of this factor, the three
times $T_{\mathrm{ct}}$, $T_{ac}$, and $T_{mc}$ are all delayed relative to
the value without IC taken into account \citep{WangLJ13}, as can be seen
from Figure \ref{fig:cooling-nu}. After $T_{mc}$, the temporal scaling
indices of $\nu _{c}$ are only slightly changed because $p$ is very close to
2.

Compared with Equation $\left( \ref{eq:PTF11agg-dyn}\right) $, the adoption
of Equation $\left( \ref{eq:dynamics-RS-work}\right) $ is to increase $\nu
_{c}$ when $t>T_{mc}$, as can be seen from Figure 2(b) of \cite{wang15}. The
consideration of IC is, on the other hand, to decrease $\nu _{c}$ by a
nearly same factor. Consequently, we find that the combination of Equation $%
\left( \ref{eq:dynamics-RS-work}\right) $ and the consideration of IC
cooling results in a cooling frequency that is very close to the one
determined in \cite{WangLJ13}, where we showed that the optical transient
PTF11agg can be accounted for by the synchrotron emission of the reverse
shock powered by a millisecond magnetar.

The decrease of $\nu _{c}$ caused by IC cooling can be easily calculated.
Substitution of $\epsilon _{e}/\epsilon _{B}\simeq 9$ and $\eta =1$ for fast
cooling into Equation $\left( \ref{eq:exactx}\right) $ gives $Y\simeq 2.5$.
During $T_{mc}<t<T_{\mathrm{dec}}$, $Y$ declines slightly, as indicated by
Equation $\left( \ref{eq:Y-T_mc-T_dec}\right) $. But because $p$ is very
close to 2, the decline is not significant so that we still have $Y>1$. When 
$t>T_{\mathrm{dec}}$, $Y$ begins to increase slowly and actually remains
close to $\sim 1.5$. Therefore, IC cooling decreases $\nu _{c}$ by a factor $%
\sim 7$.

\subsection{Optical depth to $\protect\gamma $--$\protect\gamma $ collisions}

\label{sec:gamma-gamma-collision}

The photons boosted by IC scattering acquire such a high energy that they
will annihilate with the low-energy synchrotron and thermal photons. It is
necessary to estimate the optical depth to $\gamma $--$\gamma $ collisions
so that we know under what conditions $\gamma $--$\gamma $ collisions can be
neglected. Figures \ref{fig:1e-4-spectra}--\ref{fig:gamma6-spectra} show
that the synchrotron photons are usually scattered to a maximum energy of $%
\sim 10^{13}\unit{eV}$. Therefore we set the maximum energy of the
IC-scattered photons as $h\nu _{\max }=10^{13}e_{13}\unit{eV}$, where $h$ is
the Plank constant. The frequency of the softest photons that can annihilate
with $h\nu _{\max }$ is given by \citep[e.g.,][]{Lithwick01}%
\begin{equation}
\nu _{\max ,an,\mathrm{dec}}^{\mathrm{I}}=\left( \frac{\gamma _{\mathrm{dec}%
}m_{e}c^{2}}{h}\right) ^{2}\frac{1}{\nu _{\max }}=2.8\times 10^{14}\unit{Hz}%
L_{0,47}^{3/5}M_{\text{\textrm{ej}},-4}^{-2/5}e_{13}^{-1}n^{-1/5}.
\label{eq:nv-max-an}
\end{equation}%
This frequency is in the optical to UV bands because $e_{13}\lesssim 1$. The
above calculation is carried out at time $T_{\mathrm{dec}}$ by setting the
Lorentz factor as $\gamma _{\mathrm{dec}}$. This is based on the observation
that $T_{\mathrm{dec}}\simeq 0.2\unit{days}$ at which the spectra of IC
emission are calculated below (see Section \ref{ssec:ic-anaytical} for the
reason why we choose this time).

Comparison of $\nu _{\max ,an,\mathrm{dec}}^{\mathrm{I}}$ with Equations
(10) and (11) of \cite{WangLJ13} shows that $\nu _{\max ,an,\mathrm{dec}}^{%
\mathrm{I}}$ lies in the range $\nu _{m}<\nu _{\max ,an,\mathrm{dec}}^{%
\mathrm{I}}<\nu _{c}$. At time $T_{\mathrm{dec}}$ the reverse shock is in
the slow cooling regime so that $\nu _{a}<\nu _{m}<\nu _{c}$, and the
synchrotron spectrum is $F_{\nu }=F_{\nu ,\max }\left( \nu /\nu _{m}\right)
^{-\left( p-1\right) /2}\equiv g_{1}\nu ^{-\alpha _{1}}$ for a frequency in
the range $\nu _{m}<\nu <\nu _{c}$. The synchrotron photon number with
frequency above $\nu $ is%
\begin{equation}
N_{>\nu }^{\mathrm{I}}=4\pi D_{L}^{2}\frac{g_{1}}{h}T_{\mathrm{dec}}\frac{%
\nu ^{-\alpha _{1}}}{\alpha _{1}},
\end{equation}%
where $\alpha _{1}=\left( p-1\right) /2<1$ if $p<3$.

In the center-of-mass frame of the two colliding photons, the annihilation
cross section is approximately $\sigma _{T}$ if their energy is just enough
to create electron pair. For higher energy photons the cross section goes to
zero as a power law of the photon energy. For a power-law distribution of
the seed photons, i.e. $F_{\nu }=g_{1}\nu ^{-\alpha _{1}}$, the average $%
\gamma $--$\gamma $ collision cross section can be parameterized as $\sigma
_{\gamma \gamma }=f\left( \alpha _{1}\right) \sigma _{T}$ with $f\left(
\alpha _{1}\right) <1$. In particular $f\left( 1\right) =11/180$ %
\citep{Svensson87,Lithwick01}.\footnote{%
The $\alpha _{1}$ defined in this paper is smaller by 1 than $\alpha $ in 
\cite{Lithwick01}.} For harder spectrum, i.e. smaller $\alpha _{1}$, $%
f\left( \alpha _{1}\right) $ is smaller. The synchrotron $\gamma $--$\gamma $
annihilation optical depth is 
\begin{equation}
\tau _{\mathrm{syn},\gamma \gamma }^{\mathrm{I}}=\frac{f\left( \alpha
_{1}\right) \sigma _{T}N_{>\nu _{\max ,an,\mathrm{dec}}^{\mathrm{I}}}}{4\pi
\left( 4\gamma _{\mathrm{dec}}^{2}cT_{\mathrm{dec}}\right) ^{2}}=0.97\frac{%
\left( p-2\right) ^{p-1}}{\left( p-1\right) ^{p}}n_{-1}^{7(p+1)/20}e_{13}^{%
\left( p-1\right) /2}L_{0,47}^{(7-3p)/10}\gamma _{4,4}^{p-2}\epsilon
_{e}^{p-1}\epsilon _{B,-1}^{\left( p+1\right) /4}M_{\text{\textrm{ej}}%
,-4}^{\left( p+1\right) /5}.  \label{eq:tau-gamma-gamma-dec}
\end{equation}%
The above numerical value is obtained by setting $\alpha _{1}=1$, $p=2.2$.
The actual value is slightly smaller because $\alpha _{1}<1$ and therefore $%
f\left( \alpha _{1}\right) <f\left( 1\right) $.

We see that the optical depth $\tau _{\mathrm{syn},\gamma \gamma }$ at $T_{%
\mathrm{dec}}$ is close to unity for typical parameter values. At $t<T_{%
\mathrm{dec}}$ the synchrotron radiation in the reverse shock is intense
enough that the highest energy IC photons are opaque to $\gamma $--$\gamma $
collision. But in this paper we mainly focus on the IC photons with energy $%
100\unit{GeV}$ and $1\unit{GeV}$. Equation $\left( \ref%
{eq:tau-gamma-gamma-dec}\right) $ only weakly depends on $L_{0,47}$, and $%
\gamma _{4,4}$. The parameter $\epsilon _{e}$ does not concern us because $%
\epsilon _{e}\simeq 1$. For the $100\unit{GeV}$ IC photons, a simple
extrapolation of Equation $\left( \ref{eq:tau-gamma-gamma-dec}\right) $
indicates that an ejecta mass $M_{\text{\textrm{ej}}}=7.5\times
10^{-3}M_{\odot }$ will make the IC emission attenuated at $T_{\mathrm{dec}}$%
.\ However, the ejecta as massive as $M_{\text{\textrm{ej}}}=7.5\times
10^{-3}M_{\odot }$\ are not subject to the above analysis because the above
analysis is valid only for case I, i.e. $M_{\text{\textrm{ej}}}<M_{\mathrm{ej%
},c}$\ or equivalently $T_{\mathrm{sd}}>T_{\mathrm{dec}}$, where $M_{\mathrm{%
ej},c}$\ is defined as \citep{gao13a}%
\begin{equation}
M_{\mathrm{ej},c}\sim 10^{-3}M_{\odot
}n^{1/8}I_{45}^{5/4}L_{0,47}^{-3/8}P_{0,-3}^{-5/2}\xi ^{5/4}.
\end{equation}

To find the critical ejecta mass at which the $100\unit{GeV}$\ IC photons
are completely attenuated by the soft photons, we need to extend the above
analysis to Case III, i.e. $M_{\text{\textrm{ej}}}>M_{\mathrm{ej},c}$\ or
equivalently $T_{\mathrm{sd}}<T_{\mathrm{dec}}$. In case III $T_{\mathrm{dec}%
}$\ is defined as \citep{gao13a,wang15}%
\begin{equation}
T_{\mathrm{dec}}=0.9\unit{days}L_{0,48}^{-7/3}T_{\mathrm{sd},4}^{-7/3}M_{%
\mathrm{ej},-3}^{8/3}n^{-1/3}.
\end{equation}%
$T_{\mathrm{dec}}$\ could be several days as long as $M_{\mathrm{ej}%
}>10^{-3}M_{\odot }$. We will evaluate the optical depth to $\gamma $--$%
\gamma $\ collisions at the time $T_{\mathrm{sd}}\sim 1\unit{days}$. In this
case we have the following softest photons that can annihilate $h\nu _{\max
} $

\begin{equation}
\nu _{\max ,an,\mathrm{sd}}^{\mathrm{III}}=\left( \frac{\gamma _{\mathrm{sd}%
}m_{e}c^{2}}{h}\right) ^{2}\frac{1}{\nu _{\max }}=2.0\times 10^{14}\unit{Hz}%
L_{0,48}^{2}T_{\mathrm{sd},4}^{2}M_{\text{\textrm{ej}},-3}^{-2}e_{13}^{-1}.
\end{equation}%
At the time $T_{\mathrm{sd}}$\ the reverse shock is in the fast cooling
regime, viz. $\nu _{c}<\nu _{a}<\nu _{m}$, and inspection of Equations $%
\left( 28\right) -\left( 30\right) $\ of \citep{wang15} indicates that $\nu
_{m}<\nu _{\max ,an,\mathrm{sd}}^{\mathrm{III}}$. The synchrotron spectrum
is given by $F_{\nu }=F_{\nu ,\max }\left( \nu /\nu _{m}\right)
^{-p/2}\left( \nu _{m}/\nu _{c}\right) ^{-1/2}\equiv g_{2}\nu ^{-\alpha
_{2}} $\ for $\nu >\nu _{m}$, where $\alpha _{2}=p/2$. The synchrotron
photon number with frequency above $\nu $\ is%
\begin{equation}
N_{>\nu }^{\mathrm{III}}=4\pi D_{L}^{2}\frac{g_{2}}{h}T_{\mathrm{sd}}\frac{%
\nu ^{-\alpha _{2}}}{\alpha _{2}}.
\end{equation}%
The synchrotron $\gamma $--$\gamma $\ annihilation optical depth is 
\begin{equation}
\tau _{\mathrm{syn},\gamma \gamma }^{\mathrm{III}}=\frac{f\left( \alpha
_{2}\right) \sigma _{T}N_{>\nu _{\max ,an,\mathrm{sd}}^{\mathrm{III}}}}{4\pi
\left( 4\gamma _{\mathrm{sd}}^{2}cT_{\mathrm{sd}}\right) ^{2}}=\frac{6.2}{p}%
\left( \frac{p-2}{p-1}\right) ^{p-1}e_{13}^{p/2}\gamma
_{4.4}^{p-2}L_{0,48}^{-\frac{11p}{4}-\frac{7}{2}}T_{\mathrm{sd},4}^{-\frac{7p%
}{2}-6}\epsilon _{e}^{p-1}\epsilon _{B,-1}^{\frac{p+2}{4}}M_{\mathrm{ej}%
,-3}^{3p+5}.
\end{equation}%
The above numerical value is evaluated at $p=2.2$\ and $f\left( \alpha
_{2}\right) $\ is set to $f\left( 1\right) $\ because $p$\ is close to 2 and
therefore $\alpha _{2}$\ is close to 1. For the $100\unit{GeV}$\ IC photons,
adopting $L_{0}=10^{47}\unit{erg}\unit{s}^{-1}$\ and $T_{\mathrm{sd}}=10^{5}%
\unit{s}$, we find $\tau _{\mathrm{syn},\gamma \gamma }^{\mathrm{III}}=1$\
when $M_{\mathrm{ej}}=M_{\mathrm{ej},c}^{\mathrm{IC}}\simeq 3\times
10^{-3}M_{\odot }$. Reducing $\epsilon _{B}$\ can help raise the critical
ejecta mass but its role is limited because $\tau _{\mathrm{syn},\gamma
\gamma }^{\mathrm{III}}$\ depends on $M_{\mathrm{ej}}$\ very sensitively.

Now we turn to the thermal photons. At $T_{\mathrm{dec}}$ the electrons in
the reverse shock are in slow cooling regime so that the energy is radiated
dominantly by the electrons with Lorentz factor $\gamma _{c}$. The total
synchrotron power is%
\begin{equation}
L_{\mathrm{syn}}=P_{\mathrm{syn}}\left( \gamma _{c}\right) N_{e},
\end{equation}%
where the total number of injected electrons is%
\begin{equation}
N_{e}=\frac{L_{0}t}{\gamma _{4}m_{e}c^{2}}.
\end{equation}%
The thermal photons in the reverse shock with (observed) temperature $T$
give rise to a thermal luminosity%
\begin{equation}
L_{\mathrm{th}}=4\pi R^{2}\sigma \left( T/\gamma _{\mathrm{dec}}\right)
^{4}\gamma _{\mathrm{dec}}^{2}.
\end{equation}%
Then Equation $\left( \ref{eq:Lth-Lsyn}\right) $ gives the temperature at
time $T_{\mathrm{dec}}$%
\begin{equation}
T=6.1\times 10^{4}\unit{K}\text{$\eta $}_{\mathrm{th}}^{1/4}L_{0,47}^{19/40}%
\text{$\gamma $}_{4,4}^{-1/4}M_{\mathrm{ej},-4}^{-2/5}n^{-3/40}\text{$%
\epsilon $}_{B,-1}^{-1/4},
\end{equation}%
or equivalently the frequency of the thermal photons%
\begin{equation}
\nu _{\mathrm{th}}\equiv \frac{k_{B}T}{h}\approx 1.3\times 10^{15}\unit{Hz}%
\text{$\eta $}_{\mathrm{th}}^{1/4}L_{0,47}^{19/40}\text{$\gamma $}%
_{4,4}^{-1/4}M_{\mathrm{ej},-4}^{-2/5}n^{-3/40}\text{$\epsilon $}%
_{B,-1}^{-1/4}.
\end{equation}%
One sees that this frequency is only weakly dependent on $\eta $$_{\mathrm{th%
}}$, which means that a value of $\eta $$_{\mathrm{th}}=0.1$ is essentially
equivalent to $\eta $$_{\mathrm{th}}=1$. This frequency is usually higher
than the minimum frequency $\left( \ref{eq:nv-max-an}\right) $ to annihilate
the hardest IC photons ($10\unit{TeV}$). As a result, approximately all
thermal photons, whose number density in the comoving frame is denoted by $%
n_{\mathrm{th},an}^{\prime }$, can annihilate the $10\unit{TeV}$ photons%
\begin{equation}
n_{\mathrm{th},an}^{\prime }=16\pi \zeta \left( 3\right) \left( \frac{\nu _{%
\mathrm{th}}^{\prime }}{c}\right) ^{3}=16\pi \zeta \left( 3\right) \left( 
\frac{\nu _{\mathrm{th}}}{c\gamma _{\mathrm{dec}}}\right) ^{3},
\end{equation}%
where $\zeta \left( z\right) $ is the Riemann zeta function and $\zeta
\left( 3\right) =1.20206$. Here the prime denotes the quantities in the
comoving frame. The optical depth for the thermal photons to annihilate the $%
10\unit{TeV}$ photons is%
\begin{equation}
\tau _{\mathrm{th},\gamma \gamma }\left( 10\unit{TeV}\right) =n_{\mathrm{th}%
}^{\prime }s\left( T^{\prime }\right) \sigma _{T}\Delta _{3}^{\prime
}=7.1\times 10^{3}s\,\text{$\eta $}_{\mathrm{th}}^{3/4}L_{0,47}^{1/8}\text{$%
\gamma $}_{4,4}^{-3/4}n^{-1/8}\text{$\epsilon $}_{B,-1}^{-3/4},
\end{equation}%
where we parameterize the effective annihilation cross section of the
thermal photons by $s\left( T^{\prime }\right) \sigma _{T}$ with $s\left(
T^{\prime }\right) <1$. The comoving width of Region 3 is given by %
\citep{WangLJ13}%
\begin{eqnarray}
\Delta _{3}^{\prime } &=&\frac{N_{e}}{4\pi r^{2}n_{3}}, \\
\frac{n_{3}}{n_{4}} &=&4\bar{\gamma}_{3}+3, \\
\bar{\gamma}_{3} &=&\frac{\gamma _{4}}{2}\frac{M_{\mathrm{ej}}c^{2}}{L_{0}t},
\\
n_{4} &=&\frac{L_{0}}{4\pi r^{2}\gamma _{4}^{2}m_{e}c^{2}}.
\end{eqnarray}%
This thermal annihilation optical depth is large for a reasonably small $%
\eta $$_{\mathrm{th}}\simeq 0.1$. Consequently we expect that the $10\unit{%
TeV}$ photons be attenuated significantly by the thermal photons.

For the $100\unit{GeV}$ and $1\unit{GeV}$ radiation that we focus on in this
paper, the softest photons that annihilate the high energy IC photons have a
frequency%
\begin{equation}
\nu _{\max ,an,\mathrm{dec}}=2.8\times 10^{16}\unit{Hz}L_{0,47}^{3/5}M_{%
\text{\textrm{ej}},-4}^{-2/5}e_{11}^{-1}n^{-1/5},
\end{equation}%
which is much harder than the thermal photons ($\nu _{\mathrm{th}}$). In
this case the thermal photons that can annihilate the high energy IC photons
is just the photons at the exponential tail of the thermal distribution%
\begin{equation}
n_{\mathrm{th},an}^{\prime }=\frac{8\pi }{c^{3}\gamma _{\mathrm{dec}}^{3}}%
\nu _{\mathrm{th}}\nu _{\max ,an,\mathrm{dec}}^{2}\exp \left( -\nu _{\max
,an,\mathrm{dec}}/\nu _{\mathrm{th}}\right) .
\end{equation}%
The annihilation optical depth is therefore much small, whatever value $\eta 
$$_{\mathrm{th}}$ reasonably takes,%
\begin{equation}
\tau _{\mathrm{th},\gamma \gamma }\left( 100\unit{GeV}\right) \lesssim
2.7\times 10^{-4}.
\end{equation}%
Here we suppress its parameter dependence because the most influential
factor appears in the exponent. The optical depth for $1\unit{GeV}$ photons
is even smaller.

To summarize the findings in this subsection, at time $T_{\mathrm{dec}}$,
the highest energy IC photons ($10\unit{TeV}$) are attenuated by thermal
photons. The $100\unit{GeV}$ photons, on which we focus in this paper, on
the other hand, are attenuated by synchrotron photons at time $T_{\mathrm{sd}%
}=1\unit{days}$\ if the ejecta are more massive than $3\times
10^{-3}M_{\odot }$. In short, the $\gamma $--$\gamma $ collisions have an
influence on the light curves but are usually negligible for our purpose to
calculate the $1\unit{GeV}$ and $100\unit{GeV}$ for ejecta mass $%
10^{-4}M_{\odot }$ and $10^{-3}M_{\odot }$. As a result in the following
calculations we neglect $\gamma $--$\gamma $ collisions but mention the
caveat where appropriate.

\section{Detect Newborn Magnetars by IC Emission}

\label{sec:ic}

\subsection{Inverse Compton spectra and light curves}

\label{ssec:ic-anaytical}

In this subsection we will analytically calculate the IC spectra and light
curves in order to determine the best observation strategy: observational
energy bands and cadence. We assume the Thomson limit in this analytical
treatment and refrain the consideration of the Klein-Nishina effect to
Section \ref{ssec:ic-numerical} where numerical calculations are performed.

When $t<T_{ac}$, we have $\nu _{c}<\nu _{a}<\nu _{m}$, for which case the IC
spectrum is given in Appendix \ref{sec:IC-analytic}. When $t>T_{ac}$,
whatever $\nu _{a}<\nu _{c}<\nu _{m}$ or $\nu _{a}<\nu _{m}<\nu _{c}$, we
adopt the IC spectrum calculated by \cite{Sari01}. We have discussed the
calculation of $\nu _{c}$ and $\gamma _{c}$ in Section \ref{sec:syn}. With
these results, we can easily calculate the temporal evolution of IC break
frequencies and IC flux density, which are given in Tables \ref%
{tbl:indices-nu} and \ref{tbl:indices-flux}, respectively.

\begin{figure}[tbph]
\centering\includegraphics[width=0.65\textwidth,angle=-90]{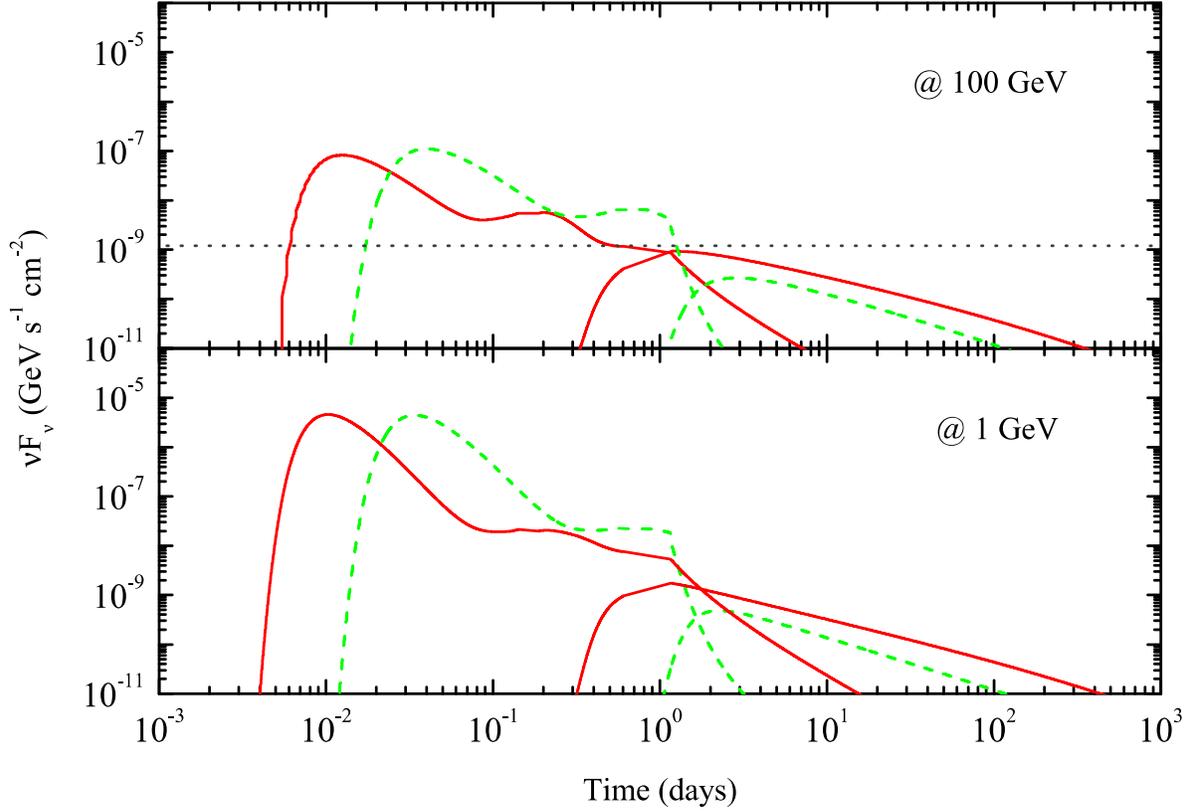}
\caption{Inverse Compton scattered high energy light curves of the reverse
shock and forward shock emission at energy bands $1\unit{GeV}$ (lower panel)
and $100\unit{GeV}$ (upper panel). The light curves for $M_{\mathrm{ej}%
}=5\times 10^{-4}M_{\odot }$ (dashed lines) are compared with the fiducial
curves for $M_{\mathrm{ej}}=10^{-4}M_{\odot }$ (solid lines). The
short-duration (long-duration) light curves are the reverse (forward) shock
emission. Other parameters are $L_{\mathrm{sd}}=10^{47}\unit{erg}\unit{s}%
^{-1}$, $T_{\mathrm{sd}}=10^{5}\unit{s}$, and the luminosity distance $%
D_{L}=10^{27}\unit{cm}$. The horizontal dotted line in the upper panel marks
the detection limit of CTA at $100\unit{GeV}$. In these calculations we
neglect $\protect\gamma $-$\protect\gamma $ collisions, which have no effect
on the light curves for the cases of $M_{\mathrm{ej}}=10^{-4}M_{\odot }$ and 
$M_{\mathrm{ej}}=5\times 10^{-4}M_{\odot }$.}
\label{fig:Mej5e-4}
\end{figure}

Provided that the observational bands lie in the range $\nu >\max \left( \nu
_{mm}^{\mathrm{IC}},\nu _{cc}^{\mathrm{IC}}\right) $, which is usually
always true if we aim in detecting $\unit{GeV}$ emission, we can read off
the evolution of IC flux density directly from Table \ref{tbl:indices-flux}.
We find that the flux density always declines except in the period $%
T_{mc}<t<T_{\mathrm{dec}}$. This fact can be clearly seen from the numerical
calculation results, i.e., Figures \ref{fig:Mej5e-4} and \ref{fig:Lsd5e47}.
But owing to the fact that $p$ is very close to 2, the light curves in the
time period $T_{mc}<t<T_{\mathrm{dec}}$ are almost flat. The rapid increase
in the flux density at the very beginning results from the ejecta becoming
progressively transparent for $\gamma $-ray emission. \cite{wang15} showed
that the opacity at X-ray and $\gamma $-ray bands is caused by the elastic
scattering of photons off free electrons in the shocked wind. Consequently
the time for the ejecta becoming transparent for $\gamma $-ray emission is
given by Equation $\left( 43\right) $ in \cite{wang15}:%
\begin{equation}
T_{\gamma ,\mathrm{thin}}=8.2\times 10^{-3}\unit{days}M_{\mathrm{ej}%
,-4}^{2/3}L_{0,47}^{-1/3}.
\end{equation}

The IC component of the reverse shock emission can be observed if it
dominates over synchrotron emission. We therefore need to determine the
frequency $\nu ^{\mathrm{IC}}$ at which the synchrotron component crosses
the IC component.

When $t<T_{ac}$, i.e. for $\nu _{c}<\nu _{a}<\nu _{m}$, Appendix \ref%
{sec:IC-analytic} shows that the IC flux density has a spectral index of $%
-1/2$ for a wide frequency range $\nu _{ca}^{\mathrm{IC}}<\nu <\nu _{mm}^{%
\mathrm{IC}}$. By assuming that $\nu ^{\mathrm{IC}}$ lies in the synchrotron
spectrum segment with spectral index $-p/2$ and between the IC break
frequencies $\nu _{ca}^{\mathrm{IC}}<\nu <\nu _{mm}^{\mathrm{IC}}$ we have%
\begin{equation}
\nu ^{\mathrm{IC}}=\left\{ 
\begin{array}{ll}
1.3\times 10^{18}\unit{Hz}\text{$\epsilon $}_{B,-1}^{1/2}\text{$\epsilon $}%
_{e}^{2}M_{\mathrm{ej},-4}^{-\frac{5-p}{2(p-1)}}\text{$\gamma $}_{4,4}^{%
\frac{2p}{p-1}}t_{3}^{\frac{4}{p-1}-\frac{3}{2}}, & t<T_{N1}; \\ 
1.5\times 10^{17}\unit{Hz}\text{$\epsilon $}_{B,-1}^{1/2}\text{$\epsilon $}%
_{e}^{2}L_{0,47}^{\frac{6}{p-1}-\frac{7}{2}}M_{\mathrm{ej},-4}^{-\frac{%
4\left( 3-p\right) }{p-1}}\text{$\gamma $}_{4,4}^{\frac{2p}{p-1}}t_{3}^{%
\frac{5\left( 3-p\right) }{p-1}}, & T_{N1}<t<T_{\mathrm{ct}}; \\ 
7.7\times 10^{16}\unit{Hz}\text{$\epsilon $}_{B,-1}^{\frac{p+1}{2(p-1)}}%
\text{$\epsilon $}_{e}^{\frac{2p-1}{p-1}}L_{0,47}^{-\frac{7p-3}{2(p-1)}}M_{%
\mathrm{ej},-4}^{\frac{2(2p-1)}{p-1}}\text{$\gamma $}_{4,4}^{\frac{2p}{p-1}%
}t_{3}^{-\frac{5p-3}{p-1}}, & T_{\mathrm{ct}}<t<T_{ac}.%
\end{array}%
\right.
\end{equation}%
We see that $\nu ^{\mathrm{IC}}$ lies in the soft X-ray bands for the
typical parameters if $t<T_{ac}$.

When $T_{ac}<t<T_{mc}$, i.e. for $\nu _{a}<\nu _{c}<\nu _{m}$, if $\nu ^{%
\mathrm{IC}}<\nu _{cc}^{\mathrm{IC}}$, we have%
\begin{equation}
\nu ^{\mathrm{IC}}=2.5\times 10^{15}\unit{Hz}\text{$\epsilon $}_{B,-1}^{-%
\frac{3(4-p)}{2(3p+2)}}\text{$\epsilon $}_{e}^{\frac{6p-13}{3p+2}}L_{0,47}^{%
\frac{7(22-3p)}{2(3p+2)}}M_{\mathrm{ej},-4}^{-\frac{12(8-p)}{3p+2}}\text{$%
\gamma $}_{4,4}^{\frac{6p}{3p+2}}t_{4}^{\frac{3(38-5p)}{3p+2}},
\end{equation}%
otherwise we have%
\begin{equation}
\nu ^{\mathrm{IC}}=3.2\times 10^{18}\unit{Hz}\text{$\epsilon $}_{B,-1}^{%
\frac{p+1}{2(p-1)}}\text{$\epsilon $}_{e}^{\frac{2p-1}{p-1}}L_{0,47}^{-\frac{%
7p-3}{2(p-1)}}M_{\mathrm{ej},-4}^{\frac{2(2p-1)}{p-1}}\text{$\gamma $}%
_{4,4}^{\frac{2p}{p-1}}t_{4}^{-\frac{5p-3}{p-1}},
\end{equation}%
if $\nu ^{\mathrm{IC}}>\nu _{cc}^{\mathrm{IC}}$.

When $t>T_{mc}$, i.e. for $\nu _{a}<\nu _{m}<\nu _{c}$, if $\nu ^{\mathrm{IC}%
}<\nu _{mm}^{\mathrm{IC}}$, we have%
\begin{equation}
\nu ^{\mathrm{IC}}=2.0\times 10^{16}\unit{Hz}\text{$\epsilon $}_{B,-1}^{%
\frac{3p^{2}-10p+4}{2(p-4)(3p+2)}}\text{$\epsilon $}_{e}^{\frac{2\left(
3p^{2}-8p-7\right) }{(p-4)(3p+2)}}L_{0,47}^{-\frac{21p^{2}-98p+176}{%
2(p-4)(3p+2)}}M_{\mathrm{ej},-4}^{\frac{12\left( p^{2}-5p+10\right) }{%
(p-4)(3p+2)}}\text{$\gamma $}_{4,4}^{\frac{2(p+2)(3p-11)}{(p-4)(3p+2)}%
}t_{4}^{\frac{15p^{2}-76p+148}{(4-p)(3p+2)}},
\end{equation}%
for $T_{mc}<t<T_{\mathrm{dec}}$;%
\begin{equation}
\nu ^{\mathrm{IC}}=2.7\times 10^{18}\unit{Hz}\text{$\epsilon $}_{B,-1}^{%
\frac{3p^{2}-10p+4}{2(p-4)(3p+2)}}\text{$\epsilon $}_{e}^{\frac{2\left(
3p^{2}-8p-7\right) }{(p-4)(3p+2)}}L_{0,47}^{-\frac{7p-34}{2(p-4)(3p+2)}}n^{%
\frac{3\left( p^{2}-5p+10\right) }{2(p-4)(3p+2)}}\text{$\gamma $}_{4,4}^{%
\frac{2(p+2)(3p-11)}{(p-4)(3p+2)}}t_{4}^{\frac{p+2}{(p-4)(3p+2)}},
\end{equation}%
for $T_{\mathrm{dec}}<t<T_{\mathrm{sd}}$;%
\begin{eqnarray}
\nu ^{\mathrm{IC}} &=&6.9\times 10^{18}\unit{Hz}\text{$\epsilon $}_{B,-1}^{%
\frac{3p^{2}-10p+4}{2(p-4)(3p+2)}}\text{$\epsilon $}_{e}^{\frac{2\left(
3p^{2}-8p-7\right) }{(p-4)(3p+2)}}L_{0,47}^{-\frac{7p-34}{2(p-4)(3p+2)}}n^{%
\frac{3\left( p^{2}-5p+10\right) }{2(p-4)(3p+2)}}  \notag \\
&&\times T_{\mathrm{sd},5}^{\frac{27p^{2}-122p+104}{16(p-4)(3p+2)}}\text{$%
\gamma $}_{4,4}^{\frac{2(p+2)(3p-11)}{(p-4)(3p+2)}}t_{5}^{-\frac{3\left(
9p^{2}-46p+24\right) }{16(p-4)(3p+2)}},
\end{eqnarray}%
for$T_{\mathrm{sd}}<t<T_{N2}$; and 
\begin{eqnarray}
\nu ^{\mathrm{IC}} &=&1.2\times 10^{17}\unit{Hz}\text{$\epsilon $}_{B,-1}^{%
\frac{3p^{2}-10p+4}{2(p-4)(3p+2)}}\text{$\epsilon $}_{e}^{\frac{2\left(
3p^{2}-8p-7\right) }{(p-4)(3p+2)}}L_{0,47}^{\frac{3p^{2}-338p+1688}{%
80(p-4)(3p+2)}}n^{\frac{117p^{2}-542p+872}{80(p-4)(3p+2)}}  \notag \\
&&\times T_{\mathrm{sd},5}^{\frac{69p^{2}-334p+424}{40(p-4)(3p+2)}}\text{$%
\gamma $}_{4,4}^{\frac{2(p+2)(3p-11)}{(p-4)(3p+2)}}t_{7}^{-\frac{3\left(
3p^{2}-18p+28\right) }{5(p-4)(3p+2)}},
\end{eqnarray}%
for $T_{N2}<t$. On the other hand, if $\nu ^{\mathrm{IC}}>\nu _{mm}^{\mathrm{%
IC}}$, we have%
\begin{equation}
\nu ^{\mathrm{IC}}=9.0\times 10^{13}\unit{Hz}\text{$\epsilon $}_{B,-1}^{%
\frac{p}{2(p-4)}}\text{$\epsilon $}_{e}^{-\frac{2(p-5)(p-1)}{p-4}}L_{0,47}^{%
\frac{4p^{2}-15p-44}{2(p-4)}}M_{\mathrm{ej},-4}^{-\frac{2\left(
p^{2}-3p-16\right) }{p-4}}\text{$\gamma $}_{4,4}^{-\frac{2(p-5)(p-2)}{p-4}%
}t_{4}^{\frac{2p^{2}-5p-40}{p-4}},
\end{equation}%
for $T_{mc}<t<T_{\mathrm{dec}}$;%
\begin{equation}
\nu ^{\mathrm{IC}}=7.9\times 10^{23}\unit{Hz}\text{$\epsilon $}_{B,-1}^{%
\frac{p}{2(p-4)}}\text{$\epsilon $}_{e}^{-\frac{2(p-5)(p-1)}{p-4}}L_{0,47}^{%
\frac{p^{2}-9p+24}{4(p-4)}}n^{-\frac{p^{2}-3p-16}{4(p-4)}}\text{$\gamma $}%
_{4,4}^{-\frac{2(p-5)(p-2)}{p-4}}t_{4}^{-\frac{(p-5)p}{2(p-4)}},
\end{equation}%
for $T_{\mathrm{dec}}<t<T_{\mathrm{sd}}$;%
\begin{equation}
\nu ^{\mathrm{IC}}=8.8\times 10^{24}\unit{Hz}\text{$\epsilon $}_{B,-1}^{%
\frac{p}{2(p-4)}}\text{$\epsilon $}_{e}^{-\frac{2(p-5)(p-1)}{p-4}}L_{0,47}^{%
\frac{p^{2}-9p+24}{4(p-4)}}n^{-\frac{p^{2}-3p-16}{4(p-4)}}\text{$\gamma $}%
_{4,4}^{-\frac{2(p-5)(p-2)}{p-4}}T_{\mathrm{sd},5}^{-\frac{8p^{2}-33p-12}{%
16(p-4)}}t_{5}^{\frac{7}{16}-\frac{1}{4-p}},
\end{equation}%
for $T_{\mathrm{sd}}<t<T_{N2}$; and%
\begin{equation}
\nu ^{\mathrm{IC}}=6.4\times 10^{20}\unit{Hz}\text{$\epsilon $}_{B,-1}^{%
\frac{p}{2(p-4)}}\text{$\epsilon $}_{e}^{-\frac{2(p-5)(p-1)}{p-4}}L_{0,47}^{%
\frac{20p^{2}-195p+588}{80(p-4)}}n^{-\frac{20p^{2}-75p-212}{80(p-4)}}\text{$%
\gamma $}_{4,4}^{-\frac{2(p-5)(p-2)}{p-4}}T_{\mathrm{sd},5}^{-\frac{%
20p^{2}-75p-84}{40(p-4)}}t_{7}^{\frac{5p-24}{5(p-4)}},
\end{equation}%
for $T_{N2}<t$. The above expressions are similar to Equation $\left(
5.1\right) $ in \cite{Sari01}.

To determine the maximum energy of IC photons, we need to calculate the
maximum Lorentz factor of the electrons in the shocked wind, which is given
by%
\begin{equation}
\gamma _{M}=\frac{3}{2}\frac{m_{e}c^{2}}{\sqrt{q^{3}B_{3}\left( 1+Y\right) }}%
,
\end{equation}%
if the IC cooling effect is taken into account. Here $B_{3}$ is the magnetic
field of Region 3, i.e., the shocked wind.%
\begin{equation}
\gamma _{M}=\left\{ 
\begin{array}{ll}
2.4\times 10^{5}M_{\mathrm{ej},-4}^{-1/4}\text{$\epsilon $}%
_{e}^{-1/4}t_{3}^{3/4}, & t<T_{N1} \\ 
1.9\times 10^{5}M_{\mathrm{ej},-4}^{-3/2}\text{$\epsilon $}%
_{e}^{-1/4}L_{0,47}^{5/4}t_{3}^{2}, & T_{N1}<t<T_{mc} \\ 
1.8\times 10^{7}L_{0,47}^{\frac{5p}{4(4-p)}}M_{\mathrm{ej},-4}^{-\frac{3p}{%
2(4-p)}}\text{$\gamma $}_{4,4}^{-\frac{p-2}{2(4-p)}}\text{$\epsilon $}%
_{B,-1}^{-\frac{p-2}{4(4-p)}}\text{$\epsilon $}_{e}^{-\frac{p-1}{2(4-p)}%
}t_{4}^{\frac{3p+2}{2(4-p)}}, & T_{mc}<t<T_{\mathrm{dec}} \\ 
1.8\times 10^{8}L_{0,47}^{-\frac{p}{16(4-p)}}n^{-\frac{3p}{16(4-p)}}\text{$%
\gamma $}_{4,4}^{-\frac{p-2}{2(4-p)}}\text{$\epsilon $}_{B,-1}^{-\frac{p-2}{%
4(4-p)}}\text{$\epsilon $}_{e}^{-\frac{p-1}{2(4-p)}}t_{4}^{\frac{8-3p}{8(4-p)%
}}, & T_{\mathrm{dec}}<t<T_{\mathrm{sd}}%
\end{array}%
\right.
\end{equation}%
The maximum energy of the IC photons is given by $\gamma \gamma
_{M}m_{e}c^{2}$, which is $\gtrsim 10\unit{TeV}$ for the most likely
observational window, i.e., $T_{mc}<t<T_{\mathrm{sd}}$.

With the above calculations, we can now design the observational strategy.
To clearly detect the IC component, viz. avoid the contamination from
synchrotron component, it is best to observe in $\gamma $-ray bands. From
Table \ref{tbl:indices-flux} and also from Figures \ref{fig:Mej5e-4} and \ref%
{fig:Lsd5e47}, it can be seen that the IC emission declines rapidly when $%
t>T_{\mathrm{sd}}$. In other words, the IC emission can be detected only in
the period $t\lesssim T_{\mathrm{sd}}\sim 1\unit{days}$ for typical
millisecond magnetars with $B_{p}=10^{15}\unit{G}$.

\subsection{Numerical calculations}

\label{ssec:ic-numerical}

To accurately calculate the high-energy spectrum of IC emission of the
reverse shock, the Klein-Nishina effect should be taken into account.
Unfortunately, the analytical treatment of this effect is complicated and
consists of several breaks \citep{Nakar09}. We therefore take a numerical
approach.

The IC volume emissivity in the full precision of Klein-Nishina scattering
cross section for an electron distribution $N\left( \gamma \right) $ is
given by \citep{Blumenthal70}%
\begin{equation}
j_{\nu }^{\mathrm{IC}}=3\sigma _{T}\int d\gamma N\left( \gamma \right)
\int_{\nu _{s,\min }}^{\infty }\frac{\nu }{4\gamma ^{2}\nu _{s}^{2}}g\left(
x,y\right) F_{\nu _{s}}d\nu _{s},
\end{equation}%
where $F_{\nu _{s}}$ is the synchrotron seed flux density, $\nu _{s,\min
}=\nu /4\gamma ^{2}$, $g\left( x,y\right) $ is given by%
\begin{equation}
g\left( x,y\right) =2y\ln y+1+y-2y^{2}+\frac{1}{2}\frac{x^{2}y^{2}}{1+xy}%
\left( 1-y\right) ,
\end{equation}%
and $x$, $y$ are defined as%
\begin{equation}
x=\frac{4\gamma h\nu _{s}}{m_{e}c^{2}},\ y=\frac{h\nu }{x\left( \gamma
m_{e}c^{2}-h\nu \right) }=\frac{\nu }{4\gamma ^{2}\nu _{s}-x\nu }.
\end{equation}

We calculate the IC emission of both the reverse shock and the forward shock
numerically, assuming the neutron star mergers locate at a luminosity
distance $D_{L}=10^{27}\unit{cm}$. In Figures \ref{fig:Mej5e-4} and \ref%
{fig:Lsd5e47} we choose the fiducial model (solid lines) as the one with
parameters $M_{\mathrm{ej}}=10^{-4}M_{\odot }$, $L_{\mathrm{sd}}=10^{47}%
\unit{erg}\unit{s}^{-1}$, and $T_{\mathrm{sd}}=10^{5}\unit{s}$. To see the
effect of varying $M_{\mathrm{ej}}$, we show the light curves for $M_{%
\mathrm{ej}}=5\times 10^{-4}M_{\odot }$ as dashed lines. It can be seen that
the more massive the ejecta, the easier for it to be observed. The figure
also shows that the IC emission from forward shock is unlikely to be
observed.

\begin{figure}[tbph]
\centering\includegraphics[width=1.0\textwidth,angle=0]{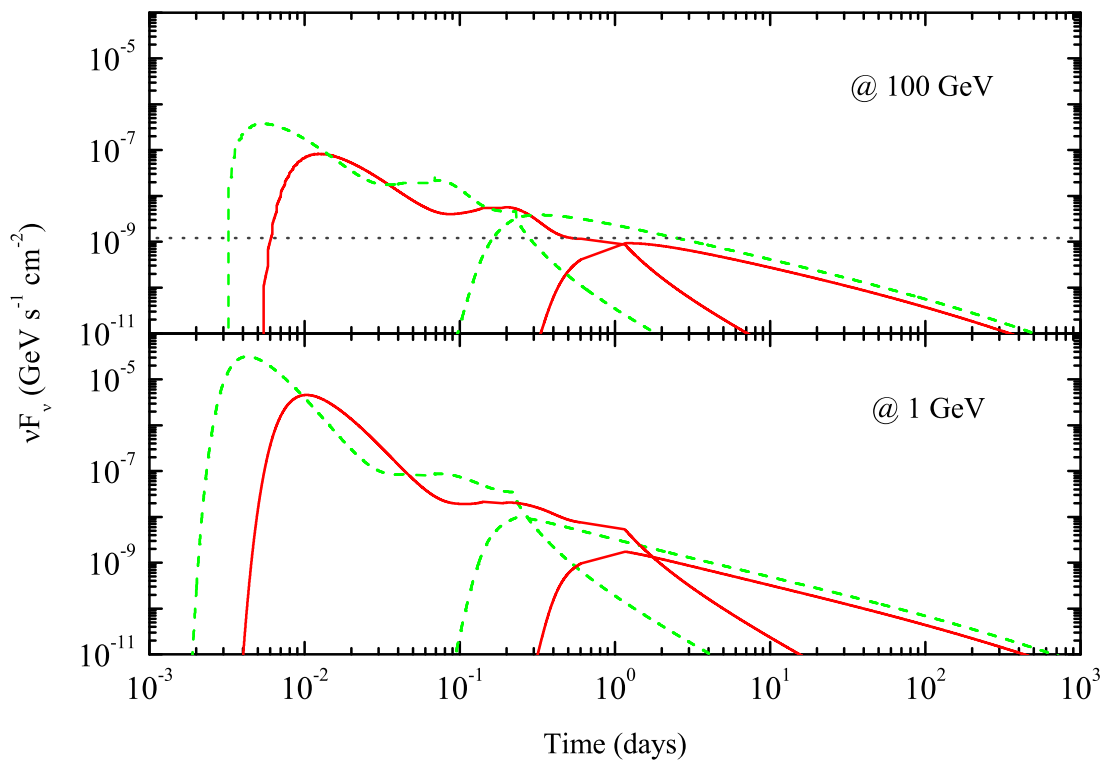}
\caption{The same as Figure \protect\ref{fig:Mej5e-4} but with $L_{\mathrm{sd%
}}=5\times 10^{47}\unit{erg}\unit{s}^{-1}$ for the dashed lines. The total
rotational energy of the magnetar is the same as Figure \protect\ref%
{fig:Mej5e-4} so $T_{\mathrm{sd}}=2\times 10^{4}\unit{s}$ (dashed lines).
The $\protect\gamma $-$\protect\gamma $ collisions have no influence on the
resulting light curves for these cases.}
\label{fig:Lsd5e47}
\end{figure}

To assess the influence of a different value of $L_{\mathrm{sd}}$, we
compare the light curves for $L_{\mathrm{sd}}=5\times 10^{47}\unit{erg}\unit{%
s}^{-1}$ with the fiducial model in Figure \ref{fig:Lsd5e47}. What should be
noted is that the rotational energy of the magnetars is fixed to be $10^{52}%
\unit{erg}$. As a result, with a spin-down power of $L_{\mathrm{sd}}=5\times
10^{47}\unit{erg}\unit{s}^{-1}$, its spin-down timescale is reduced to $T_{%
\mathrm{sd}}=2\times 10^{4}\unit{s}$. From Figure \ref{fig:Lsd5e47} we see
that the IC emission power is enhanced by the more powerful magnetars, but
its duration is shorten significantly. Consequently, only high-cadence
observations can reveal the formation of post-merger magnetars with a larger
spin-down power.

In the above calculations we set the Lorentz factor of the unshocked wind
(Region 4) as $\gamma _{4}=10^{4}$, as determined in the literature %
\citep{atoyan99,dai04,WangLJ13}. The ambient medium density is set as $n=0.1%
\unit{cm}^{-3}$ \citep{berger05,soderberg06,berger07,WangLJ13}. We will
first observe the reverse shock emission at time $\sim T_{\gamma ,\mathrm{%
thin}}$ and then the forward shock emission at time $\sim T_{\mathrm{dec}}$
if the forward shock $\gamma $ ray can be observed.

\begin{figure}[tbph]
\centering\includegraphics[width=0.6\textwidth,angle=-90]{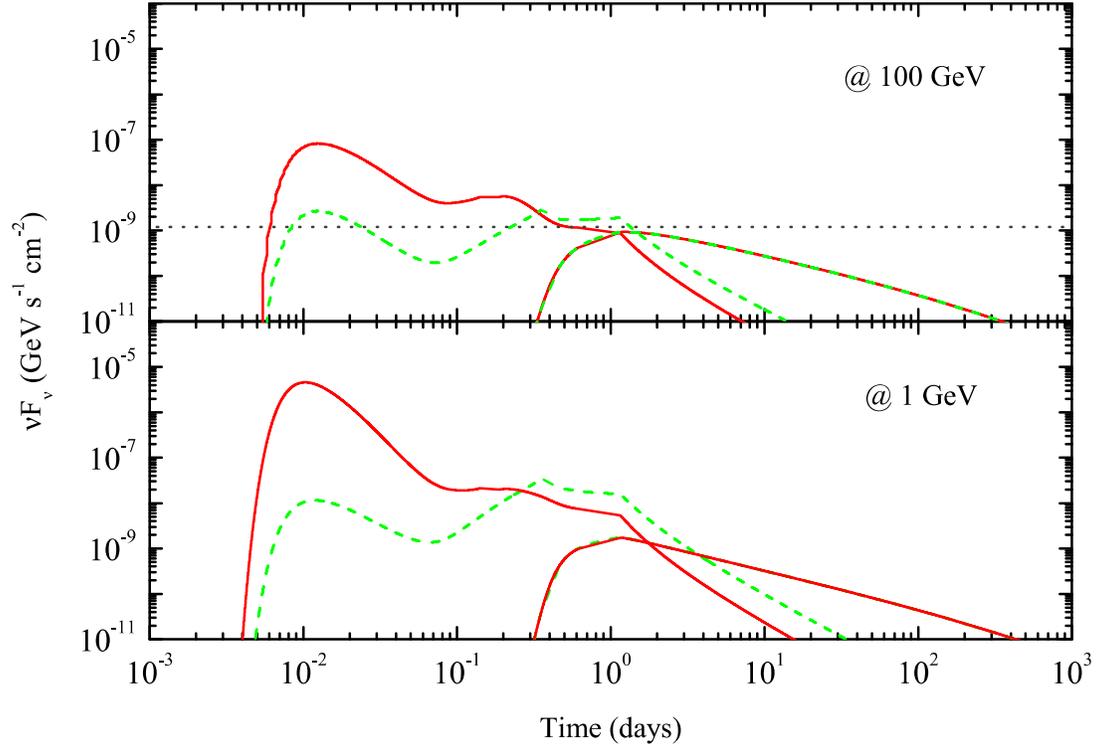}
\caption{The same as Figure \protect\ref{fig:Mej5e-4} but with $\protect%
\gamma _{4}=10^{6}$ for the dashed lines. The $\protect\gamma $-$\protect%
\gamma $ collisions have no influence on the resulting light curves for
these cases.}
\label{fig:gamma6lc}
\end{figure}

Figure \ref{fig:gamma6lc} shows the effects of varying $\gamma _{4}$ on the
resulting IC light curves. We see that increasing $\gamma _{4}$ will
suppress the early IC emission but enhance it later on. This behaviour can
be understood as follows. At beginning, the ejecta velocity is low and the
electrons' individual energy in the reverse shock is high enough to inverse
scatter the photons to high energy. As a result, the IC emission intensity
is enhanced because of the lower Lorentz factor of the unshocked wind and
hence more numerous electrons in the shocked wind (the wind power $L_{%
\mathrm{sd}}$ is fixed here). At later time, however, the ejecta become
relativistic and the electrons' individual energy in the reverse shock is
reduced. In this case, a larger Lorentz factor of the unshocked wind is
prone to produce more intense IC emission.

Here we choose the observational bands at energies of $1\unit{GeV}$ and $100%
\unit{GeV}$. The energy band at $100\unit{GeV}$ is chosen here because this
is in the sensitive region for CTA and far below the high-energy cut of the
reverse shock IC emission $\gamma \gamma _{M}m_{e}c^{2}$. We choose the
energy band at $1\unit{GeV}$ since this is the most sensitive observational
band for Fermi/LAT and it is also near the peak energy of the IC spectrum.
These points can be easily seen from Figures \ref{fig:1e-4-spectra}-\ref%
{fig:gamma6-spectra}.

To appreciate the IC spectrum and its evolution, we show the spectra at
times $t=0.2\unit{days}$ and $t=1\unit{day}$ in Figures \ref%
{fig:1e-4-spectra} (the fiducial model, $M_{\mathrm{ej}}=10^{-4}M_{\odot }$, 
$L_{\mathrm{sd}}=10^{47}\unit{erg}\unit{s}^{-1}$, and $T_{\mathrm{sd}}=10^{5}%
\unit{s}$), \ref{fig:5e-4-spectra} ($M_{\mathrm{ej}}=5\times 10^{-4}M_{\odot
}$), \ref{fig:Lsd5e47-spectra} ($L_{\mathrm{sd}}=5\times 10^{47}\unit{erg}%
\unit{s}^{-1}$), and \ref{fig:gamma6-spectra} ($\gamma _{4}=10^{6}$). From
Figure \ref{fig:Mej5e-4} we see that the second peak occurs at $t\simeq 0.2%
\unit{days}$ for the fiducial model. This is why we choose to calculate one
of the IC spectra at $t=0.2\unit{days}$. The reason we do not choose the
time when the first peak occurs is that it happens too early and is
therefore not easy to be caught by a regular-cadence observation. We choose
to calculate the other spectrum at $t=1\unit{day}$ because $T_{\mathrm{sd}%
}\simeq 1\unit{day}$ for a typical magnetar so that its emission is still
very strong to be caught by telescopes. These figures also show that the
frequencies $\nu ^{\mathrm{IC}}$ at which IC emission dominates over
synchrotron emission usually occur in the X-ray and $\gamma $-ray bands for
typical parameters.

\section{Discussions}

\label{sec:discuss}

Now our theoretical understanding of the physical processes giving rise to
particle acceleration is still incomplete. Fortunately the radiation spectra
of PWNe can be used to infer the particle acceleration processes. The
radiation spectra of PWNe from radio to a few hundred $\unit{MeV}$ is due to
synchrotron emission and the high energy part can be interpreted as inverse
Compton scattering. In this paper we assume that the shock accelerated
electrons have a power-law distribution index $2<p<3$, as is the case for
most PWNe. However there are some cases that the electrons have a flat
(broken) power law slope $N\left( E\right) \propto E^{-p}$, where $1<p<2$ %
\citep{Kargaltsev15}. Such an electron distribution slope is inconsistent
with diffusive shock acceleration (DSA). Theoretical effort suggests that
the flat electron distribution could be a result of magnetic reconnection at
the termination shock \citep{Lyubarsky03,Lyubarsky08}.

Based on the idea of driven magnetic reconnection, \cite{Lyubarsky08}
managed to show that the particle distribution can be well approximated by a
power-law with index $-1$ and an exponential cutoff. This result is
encouraging but the power-law slop is systematically flatter than
observational data. 3D simulations \citep{Sironi11b,Kargaltsev15} found that
the electrons accelerated by driven reconnection could have a flat
distribution only if the parameter $\xi =\lambda /\sigma r_{L}$, where $%
\lambda $ is the stripe wavelength, $r_{L}$ the Larmor radius of the
electrons upstream of the termination shock, takes on an unrealistic large
value. Otherwise the electrons would be in a Maxwellian distribution.
Therefore the acceleration of electrons in the relativistic wind is still an
unresolved puzzle. It is nevertheless valuable to investigate the emission
characteristics of a flat electron distribution in future work.

Another assumption of this paper is the unmagnetized upstream of the
termination shock. This assumption is plausible for a newborn millisecond
magnetar examined in this paper. We should nevertheless consider the case
where the pulsar wind is still magnetized at the termination shock, which is
the subject we will examine in an accompanying paper \citep{Liu16}.

\section{ Conclusions}

\label{sec:conclusion}

It seems that we are in an era of deciphering the central engines of SGRBs,
given that the next generation of GW detectors are already begining to
detect nearby compact binary mergers \citep{Abbott16b} and that CTA and
Fermi/LAT are becoming sensitive enough to catch the high-energy $\gamma $%
-ray signals.

In previous papers \citep{WangLJ13,wang15}, we calculated the radio and
optical/UV radiation of the reverse shock powered by the post-merger
millisecond magnetars. We find that the optical transient PTF11agg can be
neatly interpreted by the reverse shock synchrotron emission powered by a
millisecond magnetar.

In this paper we first evaluate the effects of IC cooling on the synchrotron
spectrum and light curves, which were ignored in previous papers. The IC
cooling only affects the cooling frequency $\nu _{c}$ of the synchrotron
emission and the temporal scaling indices of $\nu _{c}$ are changed only
slightly. The IC emission enhances the cooling of electrons such that $\nu
_{c}$ is reduced relative to that when IC is ignored. We find, however, that
the combination of Equation $\left( \ref{eq:dynamics-RS-work}\right) $ and
IC effect is to result in a cooling frequency that is very close to that
calculated in \cite{WangLJ13}, where the light curves of PTF11agg were
calculated.

We then further explore the high-energy emission caused by synchrotron
self-Compton (SSC) scattering powered by post-merger millisecond magnetars
after analytically estimating the attenuation caused by the $\gamma $-$%
\gamma $ collisions between high-energy IC photons and the low-energy
thermal and synchrotron photons. The SSC emission, lasting for $\sim 1$ day,
usually dominates synchrotron emission at X-ray bands and extends to the
high-energy cutoff at $\gtrsim 10\unit{TeV}$.

\begin{figure}[tbph]
\centering\includegraphics[width=1.0\textwidth,angle=0]{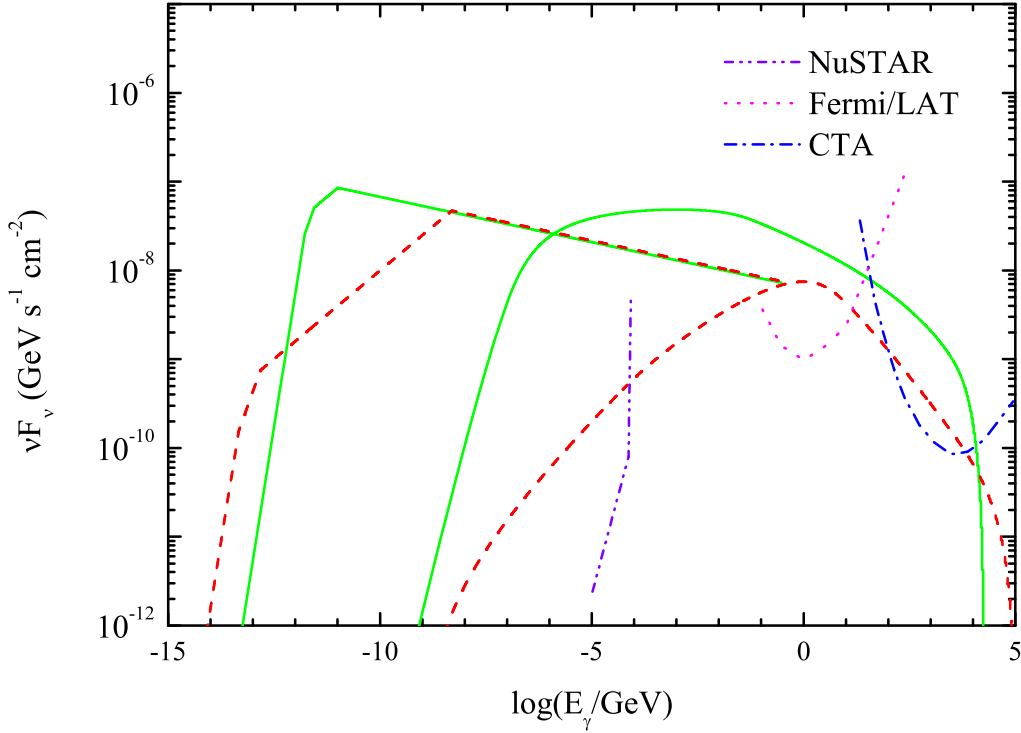}
\caption{Synchrotron (low-frequency components) and IC (high-frequency
components) spectra at $t=0.2\unit{days}$ (solid lines) and $t=1\unit{days}$
(dashed lines) for the parameters $M_{\mathrm{ej}}=10^{-4}M_{\odot }$, $L_{%
\mathrm{sd}}=10^{47}\unit{erg}\unit{s}^{-1}$, $T_{\mathrm{sd}}=10^{5}\unit{s}
$. Inclusion of $\protect\gamma $-$\protect\gamma $ collisions would make
the early emission (solid lines) at $\gtrsim 10\unit{TeV}$ dimmer. However,
given that $10\unit{TeV}$ is close to the IC high-energy cut-off, we expect
that the IC spectrum for this case is only slightly modified by including
the (thermal) $\protect\gamma $-$\protect\gamma $ collisions. $\protect%
\gamma $-$\protect\gamma $ collisions have no influence for the other case
depicted here.}
\label{fig:1e-4-spectra}
\end{figure}

\begin{figure}[tbph]
\centering\includegraphics[width=1.0\textwidth,angle=0]{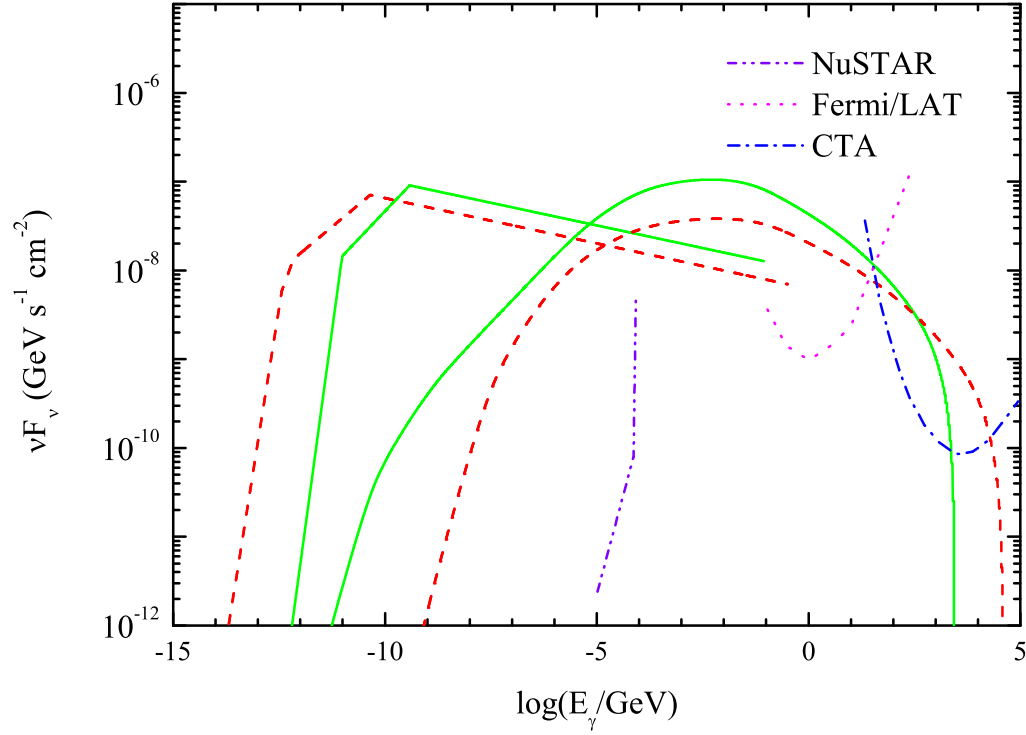}
\caption{The same as Figure \protect\ref{fig:1e-4-spectra} but with $M_{%
\mathrm{ej}}=5\times 10^{-4}M_{\odot }$, other parameters being the same. $%
\protect\gamma $-$\protect\gamma $ collisions have no influence for the
cases presented here.}
\label{fig:5e-4-spectra}
\end{figure}

\begin{figure}[tbph]
\centering\includegraphics[width=1.0\textwidth,angle=0]{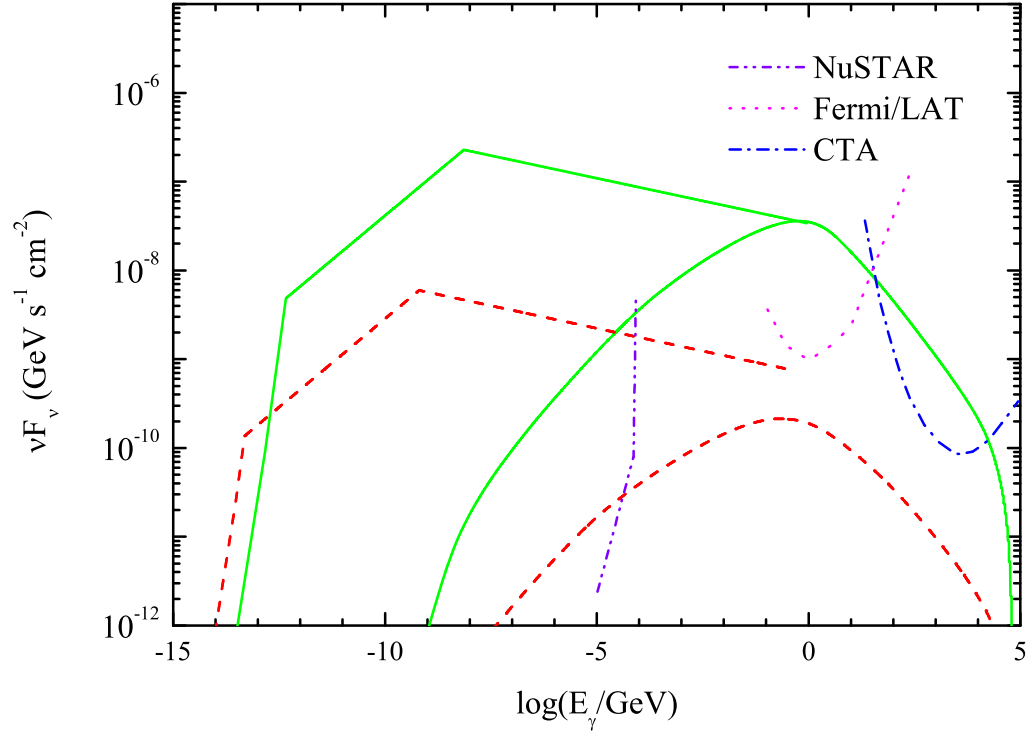}
\caption{The same as Figure \protect\ref{fig:1e-4-spectra} but with $L_{%
\mathrm{sd}}=5\times 10^{47}\unit{erg}\unit{s}^{-1}$ and therefore $T_{%
\mathrm{sd}}=2\times 10^{4}\unit{s}$, other parameters being the same. $%
\protect\gamma $-$\protect\gamma $ collisions have no effect on the spectra
for the cases presented here.}
\label{fig:Lsd5e47-spectra}
\end{figure}

\begin{figure}[tbph]
\centering\includegraphics[width=0.6\textwidth,angle=-90]{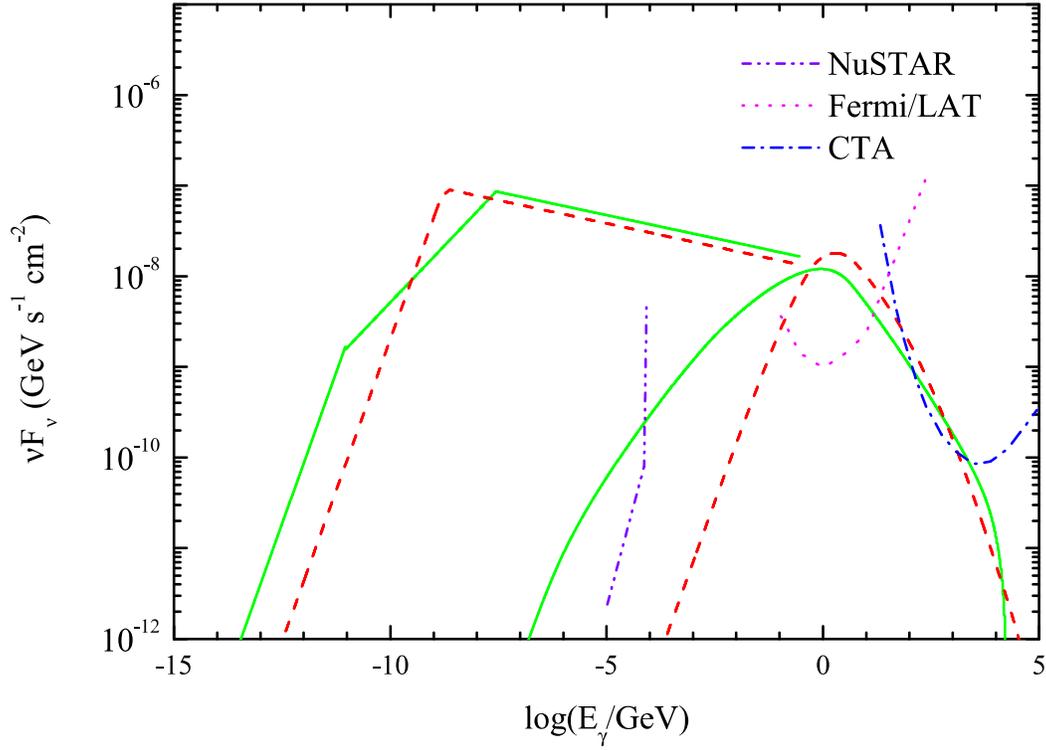}
\caption{The same as Figure \protect\ref{fig:1e-4-spectra} but with $\protect%
\gamma _{4}=10^{6}$, other parameters being the same. $\protect\gamma $-$%
\protect\gamma $ collisions have no effect on the spectra for the cases
presented here.}
\label{fig:gamma6-spectra}
\end{figure}

We find that for a typical magnetar at a distance $10^{27}\unit{cm}$, the
SSC emission can be detected by both Fermi/LAT and CTA during its whole
spin-down period $T_{\mathrm{sd}}$. Figures \ref{fig:1e-4-spectra}-\ref%
{fig:gamma6-spectra} show that the high-energy emission can be detected most
sensitively and feasibly at the energy bands $1\unit{GeV}$ and $100\unit{GeV}
$. \textit{NuSTAR} \citep{Harrison13} can also be helpful in some cases (see
Figures \ref{fig:1e-4-spectra}-\ref{fig:gamma6-spectra}) in identifying the
SSC emission in X-ray.

Comparison of Figures \ref{fig:1e-4-spectra}-\ref{fig:gamma6-spectra} shows
that the SSC component evolves more rapidly if the ejecta are less massive
or the spin-down power of the central magnetars is more powerful. The
difference of less massive ejecta and a more powerful magnetar can be
discerned by noting the fact that the more powerful the central magnetar,
the more pronounced the synchrotron component compared to the SSC component,
as can been seen by comparing the respective spectra at $t=0.2\unit{days}$
for the cases $M_{\mathrm{ej}}=10^{-4}M_{\odot }$ (Figure \ref%
{fig:1e-4-spectra}) and $L_{\mathrm{sd}}=5\times 10^{47}\unit{erg}\unit{s}%
^{-1}$ (Figure \ref{fig:Lsd5e47-spectra}). Consequently, if the observation
cadence on the same sky area is 5 times per day, the first detection of a
spectrum from optical to $\unit{GeV}$ band (at $t\sim 0.2\unit{days}$ from
the magnetar birth) contains much information about the central magnetars
and the ejecta mass. Meanwhile, we find that the high-energy IC emission is
only clearly predicted when $M_{\mathrm{ej}}\lesssim 10^{-3}M_{\odot }$.

Fermi/LAT has been in orbit for more than 6 years since its launch. The
high-energy reverse shock emission from post-merger millisecond magnetars
is, however, very rare so far \citep{Acero15}.\ This can be understood in
the following ways. If SGRBs are indeed produced by compact binary mergers
and a significant fraction of such mergers results in stable magnetars, we
should be able to observe the high-energy reverse shock emission. However,
SGRBs are usually located at redshifts $z\sim 0.5$ \citep{Berger14}, which
is beyond the detection limit, a few $10^{27}\unit{cm}$, of Fermi/LAT.

Inspection of the sensitivity of \textit{NuSTAR}, we find that \textit{NuSTAR%
} can detect the reverse shock emission powered by post-merger millisecond
magnetars up to redshifts $z\gtrsim 1$, i.e. the redshifts for most
frequently occurred SGRBs. Future improvement on the sensitivity of CTA will
extend the detection limit for high-energy reverse shock emission to $%
z\gtrsim 1$.

\begin{acknowledgements}
We would like to thank the anonymous referee for very valuable suggestions which have allowed us
to improve our manuscript significantly. This work is supported by
the National Basic Research Program (``973" Program)
of China (grant Nos. 2014CB845800 and 2013CB834900) and the 
National Natural Science Foundation of China (grant Nos. 11573014, U1331202 and 11322328). 
X.F.W. is also partially supported by the
Youth Innovation Promotion Association (2011231), and the Strategic Priority Research Program
``The Emergence of Cosmological Structures" (grant No. XDB09000000) of
the Chinese Academy of Sciences.
\end{acknowledgements}

\appendix

\section{Analytical IC spectrum in the case of $\protect\nu _{c}<\protect\nu %
_{a}<\protect\nu _{m}$}

\label{sec:IC-analytic}

In this appendix, we adopt the approximation by \cite{Sari01} to calculate
the IC spectrum for use in this paper. To our knowledge, the IC spectra of
GRB afterglows for all six cases have been analytically integrated by \cite%
{Gou07} and \cite{Gao13b}, besides \cite{Sari01}. The relevant cases for
this paper are $\nu _{c}<\nu _{a}<\nu _{m}$, $\nu _{a}<\nu _{c}<\nu _{m}$,
and $\nu _{a}<\nu _{m}<\nu _{c}$. For the later two cases we use the result
given by \cite{Sari01}.

In the case of $\nu _{c}<\nu _{a}<\nu _{m}$, \cite{Gao13b} considered the
heating of low-energy electrons due to synchrotron absorption. For our
purpose here, only the high energy IC emission concerns us and so we ignore
this low-energy heating. The synchrotron spectrum in the case of $\nu
_{c}<\nu _{a}<\nu _{m}$ is given by

\begin{equation}
F_{\nu }=\left\{ 
\begin{array}{ll}
F_{\nu ,\max }\left( \nu /\nu _{c}\right) ^{2}\left( \nu _{c}/\nu
_{a}\right) ^{3}, & \nu <\nu _{c} \\ 
F_{\nu ,\max }\left( \nu /\nu _{a}\right) ^{5/2}\left( \nu _{a}/\nu
_{c}\right) ^{-1/2}, & \nu _{c}<\nu <\nu _{a} \\ 
F_{\nu ,\max }\left( \nu /\nu _{c}\right) ^{-1/2}, & \nu _{a}<\nu <\nu _{m}
\\ 
F_{\nu ,\max }\left( \nu /\nu _{m}\right) ^{-p/2}\left( \nu _{m}/\nu
_{c}\right) ^{-1/2}, & \nu _{m}<\nu <\nu _{M}%
\end{array}%
\right.
\end{equation}%
The IC spectrum is given by the following integration \citep{Sari01}%
\begin{equation}
f_{\nu }^{\mathrm{IC}}=R\sigma _{T}\int_{\min \left( \gamma _{m},\gamma
_{c}\right) }^{\infty }d\gamma N(\gamma )\int_{0}^{x_{0}}dx\,f_{\nu _{s}}(x),
\label{eq:fc}
\end{equation}%
where $N(\gamma )$ is the electron number density per unit $\gamma $, $%
f_{\nu _{s}}(x)$ is the synchrotron seed photon flux density, $x=\nu
/4\gamma ^{2}\nu _{s}$ with $\nu $ the frequency of emitting IC photons.

In fast cooling regime, the electron distribution $N(\gamma )$ is given by
Equation $\left( 2.2\right) $ in \cite{Sari01}. If $\nu <4\gamma ^{2}\nu
_{c}x_{0}$, the inner integral in Equation $\left( \ref{eq:fc}\right) $ gives%
\begin{equation}
I_{1}\simeq \frac{5}{3}F_{\nu ,\max }x_{0}\left( \frac{\text{$\nu $}_{c}}{%
\text{$\nu $}_{a}}\right) ^{1/2}\frac{\nu }{4\gamma ^{2}\text{$\nu $}%
_{a}x_{0}}.
\end{equation}%
If $4\gamma ^{2}\nu _{c}x_{0}<\nu <4\gamma ^{2}\nu _{a}x_{0}$, this same
integral is%
\begin{equation}
I_{1}^{\prime }\simeq \frac{4}{3}F_{\nu ,\max }x_{0}\left( \dfrac{\nu _{c}}{%
\nu _{a}}\right) ^{1/2}\dfrac{\nu }{4\gamma ^{2}\nu _{a}x_{0}}.
\label{eq:I1_prime}
\end{equation}%
In the following calculation we take the value given by Equation $\left( \ref%
{eq:I1_prime}\right) $ for $\nu <4\gamma ^{2}\nu _{a}x_{0}$. We summarize
the inner integral as follows%
\begin{equation}
I=\left\{ 
\begin{tabular}{ll}
$I_{1}\simeq \frac{4}{3}F_{\nu ,\max }x_{0}\left( \dfrac{\nu _{c}}{\nu _{a}}%
\right) ^{1/2}\dfrac{\nu }{4\gamma ^{2}\nu _{a}x_{0}},$ & $\nu <4\gamma
^{2}\nu _{a}x_{0};$ \\ 
$I_{2}\simeq \frac{2}{3}F_{\nu ,\max }x_{0}\left( \dfrac{\nu }{4\gamma
^{2}\nu _{c}x_{0}}\right) ^{-1/2},$ & $4\gamma ^{2}\nu _{a}x_{0}<\nu
<4\gamma ^{2}\nu _{m}x_{0};$ \\ 
$I_{3}\simeq \dfrac{2}{p+2}F_{\nu ,\max }x_{0}\left( \dfrac{\nu _{c}}{\nu
_{m}}\right) ^{1/2}\left( \dfrac{\nu }{4\gamma ^{2}\nu _{m}x_{0}}\right)
^{-p/2},$ & $4\gamma ^{2}\nu _{m}x_{0}<\nu .$%
\end{tabular}%
\right.  \label{eq:I-nu_c.lt.nu_a.lt.nu_a}
\end{equation}

For the IC break frequencies we adopt the convenient notation used by \cite%
{Gao13b}, i.e., 
\begin{equation}
\nu _{ij}^{\mathrm{IC}}\equiv 4\gamma _{i}^{2}\nu _{j}x_{0},\qquad i=m,c,\
j=a,m,c.
\end{equation}%
As pointed out by \cite{Gao13b}, to get the right result for the double
integral $\left( \ref{eq:fc}\right) $, higher order terms of $\left( \ref%
{eq:I-nu_c.lt.nu_a.lt.nu_a}\right) $ should be taken into account. After
integration of Equation $\left( \ref{eq:fc}\right) $ we finally get%
\begin{eqnarray}
&&f_{\nu }^{\mathrm{IC}}\simeq R\sigma _{T}nF_{\nu ,\max }x_{0}
\label{F_nu-IC-c.lt.a.lt.m} \\
&&\left\{ 
\begin{tabular}{ll}
$\dfrac{4}{9}\left( \dfrac{\nu _{c}}{\nu _{a}}\right) ^{1/2}\left( \dfrac{%
\nu }{\nu _{ca}^{\mathrm{IC}}}\right) ,$ & $\nu <\nu _{ca}^{\mathrm{IC}};$
\\ 
$\dfrac{1}{3}\left( \dfrac{\nu }{\nu _{cc}^{\mathrm{IC}}}\right) ^{-1/2}%
\left[ 1+\ln \left( \dfrac{\nu }{\nu _{ca}^{\mathrm{IC}}}\right) \right] ,$
& $\nu _{ca}^{\mathrm{IC}}<\nu <\nu _{mc}^{\mathrm{IC}};$ \\ 
$\dfrac{1}{3}\left( \dfrac{\nu }{\nu _{cc}^{\mathrm{IC}}}\right) ^{-1/2}%
\left[ \dfrac{p+5}{3(p-1)}+\ln \left( \dfrac{\nu _{m}}{\nu _{a}}\right) %
\right] ,$ & $\nu _{mc}^{\mathrm{IC}}<\nu <\nu _{ma}^{\mathrm{IC}};$ \\ 
$\dfrac{1}{3}\left( \dfrac{\nu }{\nu _{cc}^{\mathrm{IC}}}\right) ^{-1/2}%
\left[ \dfrac{2(7-p)}{3(p-1)}+\ln \left( \dfrac{\nu _{mm}^{\mathrm{IC}}}{\nu 
}\right) \right] ,$ & $\nu _{ma}^{\mathrm{IC}}<\nu <\nu _{mm}^{\mathrm{IC}};$
\\ 
$\dfrac{1}{p+2}\left( \dfrac{\nu }{\nu _{mm}^{\mathrm{IC}}}\right) ^{-p/2}%
\dfrac{\nu _{c}}{\nu _{m}}\left[ \dfrac{6(p+1)}{(p-1)(p+2)}+\ln \left( 
\dfrac{\nu }{\nu _{mm}^{\mathrm{IC}}}\right) \right] ,$ & $\nu _{mm}^{%
\mathrm{IC}}<\nu .$%
\end{tabular}%
\right.  \notag
\end{eqnarray}%
The result given by \cite{Gou07} is exactly correct if we assume that all
higher-order terms in Equation $\left( \ref{eq:I-nu_c.lt.nu_a.lt.nu_a}%
\right) $ can be ignored.

\clearpage
\begin{table}[tbp]
\caption{Analytical temporal scaling indices of various parameters of the
reverse shock IC emission. }
\label{tbl:indices-nu}
\begin{center}
{\small 
\begin{tabular}{lccccccc}
\hline\hline
& $\nu _{c}$ & $\nu _{ca}^{\mathrm{IC}}$ & $\nu _{ma}^{\mathrm{IC}}$ & $\nu
_{cc}^{\mathrm{IC}}$ & $\nu _{mc}^{\mathrm{IC}}$ & $\nu _{mm}^{\mathrm{IC}}$
& $F_{\nu ,\max }^{\mathrm{IC}}$ \\ \hline
$t<T_{N1}$ & $-\frac{3}{2}$ & $-\frac{3p+14}{2\left( p+4\right) }$ & $-\frac{%
3p+14}{2\left( p+4\right) }$ & $-\frac{3}{2}$ & $-\frac{3}{2}$ & $-\frac{3}{2%
}$ & $-\frac{5}{2}$ \\ 
$T_{N1}<t<T_{\mathrm{ct}}$ & $-3$ & $-\frac{3p+14}{p+4}$ & $-\frac{5p+22}{p+4%
}$ & $-3$ & $-5$ & $-7$ & $-7$ \\ 
$T_{\mathrm{ct}}<t<T_{ac}$ & $9$ & $12-\frac{3p+2}{p+4}$ & $-\frac{5(p+2)}{%
p+4}$ & $21$ & $7$ & $-7$ & $-7$ \\ 
$T_{ac}<t<T_{mc}$ & $9$ & $\frac{6(8-3p)}{5}$ & $-\frac{2(9p+11)}{5}$ & $21$
& $7$ & $-7$ & $-7$ \\ 
$T_{mc}<t<T_{\mathrm{dec}}$ & $\frac{8+5p}{4-p}$ & $\frac{2\left(
p+10\right) }{4-p}-\frac{13}{5}$ & $-\frac{23}{5}$ & $\frac{7(p+4)}{4-p}$ & $%
\frac{7p}{4-p}$ & $-7$ & $-7$ \\ 
$T_{\mathrm{dec}}<t<T_{\mathrm{sd}}$ & $-\frac{2}{4-p}$ & $-\left( \frac{p}{%
2(4-p)}+\frac{3}{5}\right) $ & $-\frac{1}{10}$ & $-\frac{p+4}{2(4-p)}$ & $-%
\frac{p}{2(4-p)}$ & $\frac{1}{2}$ & $\frac{1}{2}$ \\ 
$T_{\mathrm{sd}}<t<T_{N2}$ & $-\left( \frac{1}{4-p}+\frac{9}{16}\right) $ & $%
-\left( \frac{1}{4-p}+\frac{3}{4}\right) $ & $-\frac{3}{4}$ & $-\left( \frac{%
2}{4-p}+\frac{9}{16}\right) $ & $-\left( \frac{1}{4-p}+\frac{9}{16}\right) $
& $-\frac{9}{16}$ & $-\frac{17}{16}$ \\ 
$T_{N2}<t$ & $-\frac{3}{5}$ & $-\frac{18}{25}$ & $-\frac{18}{25}$ & $-\frac{3%
}{5}$ & $-\frac{3}{5}$ & $-\frac{3}{5}$ & $-\frac{7}{5}$ \\ \hline
\end{tabular}%
}
\end{center}
\end{table}
\begin{table}[tbp]
\caption{Analytical temporal scaling indices of IC flux density of the
reverse shock. We have $\protect\nu _{c}<\protect\nu _{a}<\protect\nu _{m}$
if $t<T_{ac}$, $\protect\nu _{a}<\protect\nu _{c}<\protect\nu _{m}$ if $%
T_{ac}<t<T_{mc}$, and $\protect\nu _{a}<\protect\nu _{m}<\protect\nu _{c}$
if $t>T_{mc}$.}
\label{tbl:indices-flux}
\begin{center}
$%
\begin{tabular}{lcccc}
\hline\hline
$\left( \nu _{c}<\nu _{a}<\nu _{m}\right) $ & $\nu <\nu _{ca}^{\mathrm{IC}}$
& \multicolumn{2}{c}{$\nu _{ca}^{\mathrm{IC}}<\nu <\nu _{mm}^{\mathrm{IC}}$}
& $\nu _{mm}^{\mathrm{IC}}<\nu $ \\ \hline
$t<T_{N1}$ & $-\frac{2p+5}{2(p+4)}$ & \multicolumn{2}{c}{$-\frac{13}{4}$} & $%
-\frac{3p+10}{4}$ \\ 
$T_{N1}<t<T_{\mathrm{ct}}$ & $-\frac{4p+13}{p+4}$ & \multicolumn{2}{c}{$-%
\frac{17}{2}$} & $-\frac{7p+10}{2}$ \\ 
$T_{\mathrm{ct}}<t<T_{ac}$ & $-\frac{5(2p+11)}{p+4}$ & \multicolumn{2}{c}{$%
\frac{7}{2}$} & $-\frac{7(p-2)}{2}$ \\ \hline
$\left( \nu _{a}<\nu _{c}<\nu _{m}\right) $ & $\nu <\nu _{ca}^{\mathrm{IC}}$
& $\nu _{ca}^{\mathrm{IC}}<\nu <\nu _{cc}^{\mathrm{IC}}$ & $\nu _{cc}^{%
\mathrm{IC}}<\nu <\nu _{mm}^{\mathrm{IC}}$ & $\nu _{mm}^{\mathrm{IC}}<\nu $
\\ \hline
$T_{ac}<t<T_{mc}$ & $-\frac{6(17-2p)}{5}$ & $-14$ & $\frac{7}{2}$ & $-\frac{%
7(p-2)}{2}$ \\ \hline
$\left( \nu _{a}<\nu _{m}<\nu _{c}\right) $ & $\nu <\nu _{ma}^{\mathrm{IC}}$
& $\nu _{ma}^{\mathrm{IC}}<\nu <\nu _{mm}^{\mathrm{IC}}$ & $\nu _{mm}^{%
\mathrm{IC}}<\nu <\nu _{cc}^{\mathrm{IC}}$ & $\nu _{cc}^{\mathrm{IC}}<\nu $
\\ \hline
$T_{mc}<t<T_{\mathrm{dec}}$ & $-\frac{8}{5}$ & $-\frac{14}{3}$ & $-\frac{%
7\left( p+1\right) }{2}$ & $\frac{7p(p-2)}{2(4-p)}$ \\ 
$T_{\mathrm{dec}}<t<T_{\mathrm{sd}}$ & $\frac{2}{5}$ & $\frac{1}{3}$ & $%
\frac{p+1}{4}$ & $-\frac{p(p-2)}{4(4-p)}$ \\ 
$T_{\mathrm{sd}}<t<T_{N2}$ & $-\frac{3}{8}$ & $-\frac{7}{8}$ & $-\frac{9p+25%
}{32}$ & $-\left( \frac{9p+34}{32}+\frac{1}{4-p}\right) $ \\ 
$T_{N2}<t$ & $-\frac{18}{25}$ & $-\frac{6}{5}$ & $-\frac{3p+11}{10}$ & $-%
\frac{3p+14}{10}$ \\ \hline
\end{tabular}%
$%
\end{center}
\end{table}

\clearpage

\clearpage

\clearpage

\clearpage

\clearpage

\clearpage

\clearpage

\clearpage


\begin{thebibliography}{Panaitescu \& M\'{e}sz\'{a}ros(1998)}
\bibitem[Abbott et al.(2016)]{Abbott16b} Abbott, B. P., Abbott, R., Abbott,
T. D., et al. 2016, PhRvL, 116, 061102

\bibitem[Acernese et al.(2015)]{Acernese15} Acernese, F., Agathos, M.,
Agatsuma, K., et al. 2015, CQGra, 32, 024001

\bibitem[Acero et al.(2015)]{Acero15} Acero, F., Ackermann, M., Ajello, M.,
et al. 2015, ApJS, 218, 23

\bibitem[Achterberg et al.(2001)]{Achterberg01} Achterberg, A., Gallant, Y.
A., Kirk, J. G., \& Guthmann, A. W. 2001, MNRAS, 328, 393

\bibitem[Actis et al.(2011)]{Actis11} Actis, M., Agnetta, G., Aharonian, F.,
et al. (The CTA Consortium) 2011, Exp Astron, 32, 193

\bibitem[Aharonian et al.(2012)]{Aharonian12} Aharonian, F. A., Bogovalov,
S. V., \& Khangulyan, D. 2012, Nature, 482, 507

\bibitem[Arons(2012)]{Arons12} Arons, J. 2012, SSRv, 173, 341

\bibitem[Atoyan(1999)]{atoyan99} Atoyan, A. M. 1999, A\&A, 346, L49

\bibitem[Atoyan \& Aharonian(1996)]{Atoyan96} Atoyan, A. M., \& Aharonian,
F. A. 1996, MNRAS, 278, 525

\bibitem[Atwood et al.(2009)]{Atwood09} Atwood W. B., Abdo, A. A.,
Ackermann, M., et al. 2009, ApJ, 697, 1071

\bibitem[Barnes \& Kasen(2013)]{barnes13} Barnes, J. \& Kasen, D. 2013, ApJ,
775, 18

\bibitem[Barthelmy et al.(2005)]{barthelmy05} Barthelmy, S. D., Chincarini,
G., Burrows, D. N., et al. 2005, Nature, 438, 994

\bibitem[Bartos et al.(2013)]{bartos13} Bartos, I., Brady, P., \& M\'{a}rka,
S. 2013, CQGra, 30, 123001

\bibitem[Bauswein et al.(2013)]{bauswein13} Bauswein, A., Goriely, S., \&
Janka, H. T. 2013, ApJ, 773, 78

\bibitem[Begelman \& Li(1992)]{Begelman92} Begelman, M. C., \& Li, Z. Y.
1992, ApJ, 397, 187

\bibitem[Berger(2007)]{berger07} Berger, E. 2007, ApJ, 670, 1254

\bibitem[Berger(2014)]{Berger14} Berger, E. 2014, ARA\&A, 52, 43

\bibitem[Berger et al.(2013)]{berger13} Berger, E., Fong, W., \& Chornock,
R. 2013, ApJL, 774, L23

\bibitem[Berger et al.(2005)]{berger05} Berger, E., Price, P. A., Cenko, S.
B., et al. 2005, Nature, 438, 988

\bibitem[Blandford \& McKee(1976)]{blandford76} Blandford, R. D., \& McKee,
C. 1976, PhFl, 19, 1130

\bibitem[Blumenthal \& Gould(1970)]{Blumenthal70} Blumenthal, G. R. \&
Gould, R. J. 1970, RvMP, 42, 237

\bibitem[Bogovalov(1999)]{Bogovalov99} Bogovalov, S. V. 1999, A\&A, 349, 1017

\bibitem[Bucciantini et al.(2011)]{Bucciantini11} Bucciantini, N., Arons,
J., \& Amato, E. 2011, MNRAS, 410, 381

\bibitem[B\"{u}hler \& Blandford(2014)]{Buhler14} B\"{u}hler, R., \&
Blandford, R. 2014, RPPh, 77, 066901

\bibitem[Canal \& Schatzman(1976)]{Canal76} Canal, R., \& Schatzman, E.
1976, A\&A, 46, 229

\bibitem[Cenko et al.(2013)]{cenko13} Cenko, S. B., Kulkarni, S. R., Horesh,
A., Corsi, A., \& Fox, D. B. 2013, ApJ, 769, 130

\bibitem[Chatzopoulos et al.(2012)]{Chatzopoulos12} Chatzopoulos, E.,
Wheeler, J. C., \& Vinko, J. 2012, ApJ, 746, 121

\bibitem[Coroniti(1990)]{coroniti90} Coroniti, F. V. 1990, ApJ, 349, 538

\bibitem[Dai(2004)]{dai04} Dai, Z. G. 2004, ApJ, 606, 1000

\bibitem[Dai \& Liu(2012)]{Dai12} Dai, Z. G. \& Liu, R. Y. 2012, ApJ, 759, 58

\bibitem[Dai \& Lu(1998a)]{dai98a} Dai, Z. G., \& Lu, T. 1998a, A\&A, 333,
L87

\bibitem[Dai \& Lu(1998b)]{dai98b} Dai, Z. G., \& Lu, T. 1998b, PhRvL, 81,
4301

\bibitem[Dai et al.(2016)]{Dai16} Dai, Z. G., Wang, S. Q., Wang, J. S.,
Wang, L. J., \& Yu, Y. W. 2016, ApJ, 817, 132

\bibitem[Dai et al.(2006)]{dai06} Dai, Z. G., Wang, X. Y., Wu, X. F., \&
Zhang, B. 2006, Science, 311, 1127

\bibitem[Dessart et al.(2006)]{Dessart06} Dessart, L., Burrows, A., Ott, C.
D., et al. 2006, ApJ, 644, 1063

\bibitem[Eichler et al.(1989)]{eichler89} Eichler, D., Livio, M., Piran, T.,
\& Schramm, D. N. 1989, Nature, 340, 126

\bibitem[Ergma \& Tutukov(1976)]{Ergma76} Ergma, E. V., \& Tutukov, A. V.
1976, AcA, 26, 69

\bibitem[Faber \& Rasio(2012)]{rev-bin} Faber, J. A., \& Rasio, F. A. 2012,
LRR, 15, 8

\bibitem[Fang \& Zhang(2010)]{Fang10} Fang, J., \& Zhang, L. 2010, A\&A,
515, A20

\bibitem[Fern\'{a}ndez \& Metzger(2013)]{Fernandez13} Fern\'{a}ndez, R., \&
Metzger, B. D. 2013, MNRAS, 435, 502

\bibitem[Fox et al.(2005)]{fox05} Fox, D. B., Frail, D. A., Price, P. A., et
al. 2005, Nature, 437, 845

\bibitem[Fried(1959)]{Fried59} Fried, B. D. 1959, PhFl, 2, 337

\bibitem[Fryer et al.(1999)]{Fryer99} Fryer, C., Benz, W., Herant, M., \&
Colgate, S. A. 1999, ApJ, 516, 892

\bibitem[Gaensler \& Slane(2006)]{Gaensler06} Gaensler, B. M., \& Slane, P.
O. 2006, ARA\&A, 44, 17

\bibitem[Gao et al.(2015)]{Gao15} Gao, H., Ding, X., Wu, X. F., Dai, Z. G.,
\& Zhang, B. 2015, ApJ, 807, 163

\bibitem[Gao et al.(2013a)]{gao13a} Gao, H., Ding, X., Wu, X. F., Zhang, B.,
\& Dai, Z. G. 2013a, ApJ, 771, 86

\bibitem[Gao et al.(2013b)]{Gao13b} Gao, H., Lei, W. H., Wu, X. F., \&
Zhang, B. 2013b, MNRAS, 435, 2520

\bibitem[Gehrels et al.(2005)]{gehrels05} Gehrels, N., Sarazin, C. L.,
O'Brien, P. T., et al. 2005, Nature, 437, 851

\bibitem[Geng et al.(2013)]{Geng13} Geng, J. J., Wu, X. F., Huang, Y. F., \&
Yu, Y. B. 2013, ApJ, 779, 28

\bibitem[Giacomazzo \& Perna(2013)]{magnetar-sim} Giacomazzo, B., \& Perna,
R. 2013, ApJL, 771, L26

\bibitem[Giacomazzo et al.(2013)]{Giacomazzo-etal13} Giacomazzo, B., Perna,
R., Rezzolla, L., et al. 2013, ApJL, 762, L18

\bibitem[Giacomazzo et al.(2015)]{Giacomazzo15} Giacomazzo, B., Zrake, J.,
Duffell, P. C., et al. 2015, ApJ, 809, 39

\bibitem[Gompertz et al.(2014)]{Gompertz14} Gompertz, B. P., O'Brien, P. T.,
\& Wynn, G. A. 2014, MNRAS, 438, 240

\bibitem[Goriely et al.(2011)]{goriely11} Goriely, S., Bauswein, A., \&
Janka, H. T. 2011, ApJL, 738, L32

\bibitem[Gou et al.(2007)]{Gou07} Gou, L. J., Fox, D. B., \& M\'{e}sz\'{a}%
ros, P. 2007, ApJ, 668, 1083

\bibitem[Greiner et al.(2015)]{Greiner15} Greiner, J., Mazzali, P. A., Kann,
D. A., et al. 2015, Nature, 523, 189

\bibitem[Harrison et al.(2013)]{Harrison13} Harrison, F. A., Craig, W. W.,
Christensen, F. E., et al. 2013, ApJ, 770, 103

\bibitem[Harrison et al.(2001)]{Harrison01} Harrison, F. A., Yost, S. A.,
Sari, R., et al. 2001, ApJ, 559, 123

\bibitem[Harry(2010)]{Harry10} Harry, G. M. LIGO Scientific Collaboration,
2010, CQGra, 27, 084006

\bibitem[Hester(2008)]{Hester08} Hester, J. J. 2008, ARA\&A, 46, 127

\bibitem[Hoshino \& Lyubarsky(2012)]{Hoshino12} Hoshino, M., \& Lyubarsky,
Y. 2012, SSRv, 173, 521

\bibitem[Hotokezaka et al.(2013)]{hotokezaka13} Hotokezaka, K., Kiuchi, K.,
Kyutoku, K., et al. 2013, PhRvD, 87, 024001

\bibitem[Inserra et al.(2013)]{Inserra13} Inserra, C., Smartt, S. J.,
Jerkstrand, A., et al. 2013, ApJ, 770, 128

\bibitem[Just et al.(2015)]{Just15} Just, O., Bauswein, A., Pulpillo, R. A.,
Goriely, S., \& Janka, H.-T. 2015, MNRAS, 448, 541

\bibitem[Kargaltsev et al.(2015)]{Kargaltsev15} Kargaltsev, O., Cerutti, B.,
Lyubarsky, Y., \& Striani, E. 2015, SSRv, 191, 391

\bibitem[Kasen et al.(2013)]{kasen13} Kasen, D., Badnell, N. R., \& Barnes,
J. 2013, ApJ, 774, 25

\bibitem[Kasen \& Bildsten(2010)]{Kasen10} Kasen, D., \& Bildsten, L. 2010,
ApJ, 717, 245

\bibitem[Kasen et al.(2015)]{Kasen15} Kasen, D., Fern\'{a}ndez, \& Metzger,
B. D. 2015, MNRAS, 450, 1777

\bibitem[Kawanaka et al.(2013)]{Kawanaka13} Kawanaka, N., Mineshige, S., \&
Piran, T. 2013, ApJL, 777, L15

\bibitem[Kennel \& Coroniti(1984a)]{Kennel84a} Kennel, C. F., \& Coroniti,
F. V. 1984a, ApJ, 283, 694

\bibitem[Kennel \& Coroniti(1984b)]{Kennel84b} Kennel, C. F., \& Coroniti,
F. V. 1984b, ApJ, 283, 710

\bibitem[Kirk \& Skj\ae aasen(2003)]{Kirk03} Kirk, J. G., \& Skj\ae aasen,
O. 2003, ApJ, 591, 366

\bibitem[Kisaka et al.(2016)]{Kisaka16} Kisaka, S., Ioka, K., \& Nakar, E.
2016, ApJ, 818, 104

\bibitem[Kisaka et al.(2015)]{Kisaka15} Kisaka, S., Ioka, K., \& Takami, H.
2015, ApJ, 802, 119

\bibitem[Kohri \& Mineshige(2002)]{Kohri02} Kohri, K., \& Mineshige, S.
2002, ApJ, 577, 311

\bibitem[Komissarov(2013)]{Komissarov13} Komissarov, S. S. 2013, MNRAS, 428,
2459

\bibitem[Kulkarni(2005)]{kulkarni05} Kulkarni, S. R. 2005,
arXiv:astro-ph/0510256

\bibitem[Lee et al.(2009)]{Lee09} Lee, W. H., Ramirez-Ruiz, E., \& L\'{o}%
pez-C\'{a}mara, D. 2009, ApJL, 699, L93

\bibitem[Li \& Paczy\'{n}ski(1998)]{li98} Li, L. X., \& Paczy\'{n}ski, B.
1998, ApJL, 507, L59

\bibitem[Li \& Yu(2016)]{Li16} Li, S. Z., \& Yu, Y. W. 2016, ApJ, 819, 120

\bibitem[Lippuner \& Roberts(2015)]{Lippuner15} Lippuner, J., \& Roberts, L.
F. 2015, ApJ, 815, 82

\bibitem[Lithwick \& Sari(2001)]{Lithwick01} Lithwick, Y., \& Sari, R. 2001,
ApJ, 555, 540

\bibitem[Liu et al.(2016)]{Liu16} Liu, L. D., Wang, L. J., \& Dai, Z. G.
2016, submitted

\bibitem[Liu et al.(2013)]{Liu13} Liu, R. Y., Wang, X. Y., \& Wu, X. F.
2013, ApJL, 773, L20

\bibitem[Liu et al.(2015)]{Liu15} Liu, T., Gu, W. M., Kawanaka, N., \& Li,
A. 2015, ApJ, 805, 37

\bibitem[Liu et al.(2007)]{Liu07} Liu, T., Gu, W. M., Xue, L., \& Lu, J. F.
2007, ApJ, 661, 1025

\bibitem[Lyubarsky(2003)]{Lyubarsky03} Lyubarsky, Y. E. 2003, MNRAS, 345, 153

\bibitem[Lyubarsky(2005)]{Lyubarsky05} Lyubarsky, Y. 2005, AdSpR, 35, 1112

\bibitem[Lyubarsky(2010a)]{Lyubarsky10a} Lyubarsky, Y. 2010a, ApJL, 725, L234

\bibitem[Lyubarsky(2010b)]{Lyubarsky10b} Lyubarsky, Y. E. 2010b, MNRAS, 402,
353

\bibitem[Lyubarsky \& Kirk(2001)]{Lyubarsky01} Lyubarsky, Y., \& Kirk, J. G.
2001, ApJ, 547, 437

\bibitem[Lyubarsky \& Liverts(2008)]{Lyubarsky08} Lyubarsky, Y., \& Liverts,
M. 2008, ApJ, 682, 1436

\bibitem[Medvedev \& Loeb(1999)]{Medvedev99} Medvedev, M. V., \& Loeb, A.
1999, ApJ, 526, 697

\bibitem[Metzger(2012)]{Metzger12} Metzger, B. D. 2012, MNRAS, 419, 827

\bibitem[Metzger \& Berger(2012)]{MetzgerBerger12} Metzger, B. D., \&
Berger, E. 2012, ApJ, 746, 48

\bibitem[Metzger \& Fern\'{a}ndez(2014)]{MetzgerFernandez14} Metzger, B. D.,
\& Fern\'{a}ndez, R. 2014, MNRAS, 441, 3444

\bibitem[Metzger et al.(2010)]{metzger10} Metzger, B. D., Mart\'{\i}%
nez-Pinedo, G., Darbha, S., et al. 2010, MNRAS, 406, 2650

\bibitem[Metzger \& Piro(2014)]{Metzger14} Metzger, B. D., \& Piro, A. L.
2014, MNRAS, 439, 3916

\bibitem[Metzger et al.(2008)]{Metzger08} Metzger, B. D., Piro, A. L., \&
Quataert, E. 2008, MNRAS, 390, 781

\bibitem[Metzger et al.(2009)]{Metzger09} Metzger, B. D., Piro, A. L., \&
Quataert, E. 2009, MNRAS, 396, 304

\bibitem[Michel(1973)]{Michel73} Michel, F. C. 1973, ApJL, 180, L133

\bibitem[Michel(1982)]{Michel82} Michel, F. C. 1982, RvMP, 54, 1

\bibitem[Michel(1994)]{michel94} Michel, F. C. 1994, ApJ, 431, 397

\bibitem[Michel \& Li(1999)]{Michel99} Michel, F. C., \& Li, H. 1999, PhR,
318, 227

\bibitem[Mizuno et al.(2011)]{Mizuno11} Mizuno, Y., Lyubarsky, Y.,
Nishikawa, K. I., \& Hardee, P. E. 2011, ApJ, 728, 90

\bibitem[Nakar et al.(2009)]{Nakar09} Nakar, E., Ando, S., \& Sari, R. 2009,
ApJ, 703, 675

\bibitem[Nakar \& Piran(2011)]{nakar11} Nakar, E., \& Piran, T. 2011,
Nature, 478, 82

\bibitem[Narayan et al.(2001)]{Narayan01} Narayan, R., Piran, T., \& Kumar,
P. 2001, ApJ, 557, 949

\bibitem[Nicholl et al.(2014)]{Nicholl14} Nicholl, M., Smartt, S. J.,
Jerkstrand, A., et al. 2014, MNRAS, 444, 2096

\bibitem[Nomoto(1982)]{Nomoto82} Nomoto, K. 1982, ApJ, 253, 798

\bibitem[Nomoto \& Kondo(1991)]{Nomoto91} Nomoto, K., \& Kondo, Y. 1991,
ApJL, 367, L19

\bibitem[Nomoto et al.(1984)]{Nomoto84} Nomoto, K., Thielemann, F. K., \&
Yokoi, K. 1984, ApJ, 286, 644

\bibitem[Olmi et al.(2014)]{Olmi14} Olmi, B., Del Zanna, L., Amato, E.,
Bandiera, R., \& Bucciantini, N. 2014, MNRAS, 438, 1518

\bibitem[Paczy\'{n}ski(1986)]{paczynski86} Paczy\'{n}ski, B. 1986, ApJL,
308, L43

\bibitem[Palenzuela et al.(2013)]{palenzuela13} Palenzuela, C., Lehner, L.,
Ponce, M., et al. 2013, PhRvL, 111, 061105

\bibitem[Panaitescu \& Kumar(2000)]{Panaitescu00} Panaitescu, A., \& Kumar,
P. 2000, ApJ, 543, 66

\bibitem[Panaitescu \& M\'{e}sz\'{a}ros(1998)]{Panaitescu98} Panaitescu, A.,
\& M\'{e}sz\'{a}ros, P. 1998, ApJ, 501, 772

\bibitem[Papadopoulos et al.(2015)]{Papadopoulos15} Papadopoulos, A.,
D'Andrea, C. B., Sullivan, M., et al. 2015, MNRAS, 449, 1215

\bibitem[P\'{e}tri \& Lyubarsky(2007)]{Petri07} P\'{e}tri, J., \& Lyubarsky,
Y. 2007, A\&A, 473, 683

\bibitem[P\'{e}tri \& Lyubarsky(2008)]{Petri08} P\'{e}tri, J., \& Lyubarsky,
Y. 2008, IJMPD, 17, 1961

\bibitem[Piran et al.(2013)]{piran13} Piran, T., Nakar, E., \& Rosswog, S.
2013, MNRAS, 430, 2121

\bibitem[Popham et al.(1999)]{Popham99} Popham, R., Woosley, S. E., \&
Fryer, C. 1999, ApJ, 518, 356

\bibitem[Porth et al.(2014)]{Porth14} Porth, O., Komissarov, S. S., \&
Keppens, R. 2014, MNRAS, 438, 278

\bibitem[Potekhin et al.(2015)]{Potekhin15} Potekhin, A. Y., De Luca, A., \&
Pons, J. A. 2015, SSRv, 191, 171

\bibitem[Rees \& Gunn(1974)]{Rees74} Rees, M. J., \& Gunn, J. E. 1974,
MNRAS, 167, 1

\bibitem[Rezzolla et al.(2010)]{rezzolla10} Rezzolla, L., Baiotti, L.,
Giacomazzo, B., et al. 2010, CQGra, 27, 114105

\bibitem[Rezzolla et al.(2011)]{rezzolla11} Rezzolla, L., Giacomazzo, B.,
Baiotti, L., et al. 2011, ApJL, 732, L6

\bibitem[Roberts et al.(2011)]{roberts11} Roberts, L. F., Kasen, D., Lee, W.
H., \& Ramirez-Ruiz, E. 2011, ApJL, 736, L21

\bibitem[Rosswog(2005)]{rosswog05} Rosswog, S. 2005, ApJ, 634, 1202

\bibitem[Rosswog et al.(2014)]{rosswog14} Rosswog, S., Korobkin, O.,
Arcones, A., Thielemann, F. K., \& Piran, T. 2014, MNRAS, 439, 744

\bibitem[Rosswog et al.(2013)]{rosswog13} Rosswog, S., Piran, T., \& Nakar,
E. 2013, MNRAS, 430, 2585

\bibitem[Rowlinson et al.(2013)]{rowlinson13} Rowlinson, A., O'Brien, P. T.,
Metzger, B. D., et al. 2013, MNRAS, 430, 1061

\bibitem[Rowlinson et al.(2010)]{rowlinson10} Rowlinson, A., O'Brien, P. T.,
Tanvir, N. R., et al. 2010, MNRAS, 409, 531

\bibitem[Sari \& Esin(2001)]{Sari01} Sari, R, \& Esin, A. A. 2001, ApJ, 548,
787

\bibitem[Sari et al.(1998)]{Sari98} Sari, R., Piran, T., \& Narayan, R.
1998, ApJL, 497, L17

\bibitem[Siegel \& Ciolfi(2016a)]{Siegel16a} Siegel, D. M., \& Ciolfi, R.
2016a, ApJ, 819, 14

\bibitem[Siegel \& Ciolfi(2016b)]{Siegel16b} Siegel, D. M., \& Ciolfi, R.
2016b, ApJ, 819, 15

\bibitem[Sironi \& Spitkovsky(2009)]{Sironi09} Sironi, L., \& Spitkovsky, A.
2009, ApJL, 707, L92

\bibitem[Sironi \& Spitkovsky(2011a)]{Sironi11a} Sironi, L., \& Spitkovsky,
A. 2011a, ApJ, 726, 75

\bibitem[Sironi \& Spitkovsky(2011b)]{Sironi11b} Sironi, L., \& Spitkovsky,
A. 2011b, ApJ, 741, 39

\bibitem[Soderberg et al.(2006)]{soderberg06} Soderberg, A. M., Berger, E.,
Kasliwal, M., et al. 2006, ApJ, 650, 261

\bibitem[Somiya(2012)]{Somiya12} Somiya, K. (The KAGRA Collaboration) 2012,
CQGra, 29, 124007

\bibitem[Song et al.(2016)]{Song16} Song, C. Y., Liu, T., Gu, W. M., Tian,
J. X. 2016, MNRAS, 458, 1921

\bibitem[{Svensson}(1987)]{Svensson87} {Svensson}, R. 1987, MNRAS, 227, 403

\bibitem[Takami et al.(2014)]{takami14} Takami, H., Nozawa, T., \& Ioka, K.
2014, ApJL, 789, L6

\bibitem[Tanaka \& Takahara(2011)]{Tanaka11} Tanaka, S. J., \& Takahara, F.
2011, ApJ, 741, 40

\bibitem[Tanvir et al.(2013)]{tanvir13} Tanvir, N. R., Levan, A. J.,
Fruchter, A. S., et al. 2013, Nature, 500, 547

\bibitem[Tchekhovskoy et al.(2013)]{Tchekhovskoy13} Tchekhovskoy, A.,
Spitkovsky, A., Li, J. G. 2013, MNRAS, 435, L1

\bibitem[Volpi et al.(2008)]{Volpi08} Volpi, D., Del Zanna, L., Amato, E.,
\& Bucciantini, N. 2008, A\&A, 485, 337

\bibitem[Wang \& Dai(2013a)]{WangFY13} Wang, F. Y., \& Dai, Z. G. 2013a,
NatPh, 9, 465

\bibitem[Wang \& Dai(2013b)]{WangK13} Wang, K., \& Dai, Z. G. 2013b, ApJ,
772, 152

\bibitem[Wang \& Dai(2013c)]{WangLJ13} Wang, L. J., \& Dai, Z. G. 2013c,
ApJL, 774, L33

\bibitem[Wang et al.(2015a)]{wang15} Wang, L. J., Dai, Z. G., \& Yu, Y. W.
2015a, ApJ, 800, 79

\bibitem[Wang et al.(2016)]{Wang16} Wang, L. J., Wang, S. Q., Dai, Z. G., et
al. 2016, ApJ accepted, arXiv:1602.06190

\bibitem[Wang et al.(2015b)]{WangLiu15} Wang, S. Q., Liu, L. D., Dai, Z. G.,
Wang, L. J., \& Wu, X. F. 2015b, arXiv:1509.05543

\bibitem[Wang et al.(2015c)]{Wang15b} Wang, S. Q., Wang, L. J., Dai, Z. G.,
\& Wu, X. F. 2015c, ApJ, 799, 107

\bibitem[Wang et al.(2015d)]{Wang15c} Wang, S. Q., Wang, L. J., Dai, Z. G.,
\& Wu, X. F. 2015d, ApJ, 807, 147

\bibitem[Wang et al.(2001a)]{Wang01a} Wang, X. Y., Dai, Z. G., \& Lu, T.
2001a, ApJL, 546, L33

\bibitem[Wang et al.(2001b)]{Wang01b} Wang, X. Y., Dai, Z. G., \& Lu, T.
2001b, ApJ, 556, 1010

\bibitem[Wang et al.(2010)]{Wang10} Wang, X. Y., He, H. N., Li, Z., Wu, X.
F., \& Dai, Z. G. 2010, ApJ, 712, 1232

\bibitem[Wei \& Lu(1998)]{Wei98} Wei, D. M., \& Lu, T. 1998, ApJ, 505, 252

\bibitem[Weibel(1959)]{Weibel59} Weibel, E. S. 1959, PhRvL, 2, 83

\bibitem[Woosley(2010)]{Woosley10} Woosley, S. E. 2010, ApJL, 719, L204

\bibitem[Wu et al.(2014)]{Wu14} Wu, X. F., Gao, H., Ding, X., Zhang, B.,
Dai, Z. G., Wei, J. Y. 2014, ApJL, 781, L10

\bibitem[Wu et al.(2013)]{Wu13} Wu, X. F., Hou, S. J., \& Lei, W. H. 2013,
ApJL, 767, L36

\bibitem[Xue et al.(2013)]{Xue13} Xue, L., Liu, T., Gu, W. M., \& Lu, J. F.
2013, ApJS, 207, 23

\bibitem[Yu et al.(2015a)]{Yu15a} Yu, Y. B., Wu, X. F., Huang, Y. F., et al.
2015a, MNRAS, 446, 3642

\bibitem[Yu \& Dai(2007)]{Yu07} Yu, Y. W., \& Dai, Z. G. 2007, A\&A, 470, 119

\bibitem[Yu et al.(2015b)]{Yu15b} Yu, Y. W., Li, S. Z., \& Dai, Z. G. 2015b,
ApJL, 806, L6

\bibitem[Yu et al.(2013)]{Yu13} Yu, Y. W., Zhang, B., \& Gao, H. 2013, ApJL,
776, L40

\bibitem[Zhang(2013)]{zhang13} Zhang, B. 2013, ApJL, 763, L22

\bibitem[Zhang \& Kobayashi(2005)]{Zhang05} Zhang, B., \& Kobayashi, S.
2005, ApJ, 628, 315

\bibitem[Zhang \& M\'{e}sz\'{a}ros(2001)]{Zhang01} Zhang, B., \& M\'{e}sz%
\'{a}ros, P. 2001, ApJL, 552, L35

\bibitem[Zhang \& Dai(2008)]{ZhangD08} Zhang, D., \& Dai, Z. G. 2008, ApJ,
683, 329

\bibitem[Zhang \& Dai(2009)]{ZhangD09} Zhang, D., \& Dai, Z. G. 2009, ApJ,
703, 461

\bibitem[Zhang \& Dai(2010)]{ZhangD10} Zhang, D., \& Dai, Z. G. 2010, ApJ,
718, 841
\end{thebibliography}
\end{document}